\documentclass[a4paper,11pt]{article}
\usepackage{jheppub} 
\usepackage[T1]{fontenc} 

\usepackage{xfrac}

\usepackage[makeroom]{cancel}

\def\be{\begin{equation}}
\def\ee{\end{equation}}
\def\bea{\begin{eqnarray}}
\def\eea{\end{eqnarray}}
\def\beq{\begin{eqnarray}}
\def\eeq{\end{eqnarray}}
\def\bas{\begin{subequations}\begin{eqnarray}}
\def\eas{\end{eqnarray}\end{subequations}}
\def\nn{\nonumber}

\def\eps{\varepsilon}

\def\la{\langle}
\def\ra{\rangle}

\def\f{\frac}

\def\SU{\text{SU}}
\def\su{\text{su}}
\def\SO{\text{SO}}

\def\SL{\text{SL}}
\def\su{\mathfrak{su}}

\def\sl{\mathfrak{sl}}

\newcommand{\C}{{\mathbb C}}
\newcommand{\N}{{\mathbb N}}
\newcommand{\R}{{\mathbb R}}
\newcommand{\Z}{{\mathbb Z}}

\newcommand{\cE}{{\mathcal E}}

\newcommand{\cG}{{\mathcal G}}

\newcommand{\cJ}{{\mathcal J}}

\newcommand{\cL}{{\mathcal L}}
\newcommand{\cH}{{\mathcal H}}

\newcommand{\cO}{{\mathcal O}}

\newcommand{\cT}{{\mathcal T}}

\newcommand{\cD}{{\mathcal D}}
\newcommand{\cC}{{\mathcal C}}
\newcommand{\cS}{{\mathcal S}}

\newcommand{\mat} [2] {\left ( \begin{array}{#1}#2\end{array} \right ) }

\def\veta{\vec{\eta}}

\def\pp{\partial}
\def\rd{\textrm{d}}

\def\ka{\kappa}
\def\vphi{\varphi}
\def\eps{\epsilon}

\newcommand{\id}{\mathbb{I}}

\def\vlambda{\vec{\lambda}}

\def\la{\langle}
\def\ra{\rangle}

\newcommand{\bes}{\begin{eqnarray}}
\newcommand{\ees}{\end{eqnarray}}

\renewcommand{\sl}{{\mathfrak{sl}}}

\def\nn{\nonumber}
\def\pp{\partial}

\def\ka{\kappa}
\def\vphi{\varphi}
\def\eps{\epsilon}

\def\vcJ{\vec{\cJ}}

\def\hv{\widehat{v}}

\def\hA{\widehat{A}}
\def\what{\widehat}

\def\hk{\hat{k}}
\def\hj{\hat{j}}

\def\hcC{\widehat{\cC}}
\def\hcH{\widehat{\cH}}
\def\hcO{\widehat{\cO}}

\def\tN{\tilde{N}}

\def\tt{\tilde{t}}

\def\hcC{\widehat{\cC}}
\def\hcH{\widehat{\cH}}

\def\ads{AdS${}_{2}$}

\def\hcH{\widehat{\cH}}
\def\hcC{\widehat{\cC}}

\newcommand{\dt}[1]{#1'}

\def\ttau{\tilde{\tau}}
\def\ta{\tilde{a}}
\def\tphi{\tilde{\phi}}

\title{\boldmath Cosmology as a $\text{CFT}_1$ }

 \author[a]{Jibril Ben Achour,}
\author[b]{Etera R. Livine,}

\affiliation[a]{Center for Gravitational Physics, Yukawa Institute for Theoretical Physics, Kyoto University, 606-8502, Kyoto, Japan}
\affiliation[b]{ENS de Lyon, Universit\'e Lyon 1, CNRS, Laboratoire de Physique, F-69342 Lyon, France}

\abstract{
We show that the simplest FLRW cosmological system consisting in the homogeneous and isotropic massless Einstein-Scalar system
enjoys a hidden conformal symmetry under the 1D conformal group $\SL(2,\mathbb{R})$ acting as Mobius transformations in proper time. This invariance is made explicit through the mapping of FLRW cosmology onto conformal mechanics. On the one hand, we identify the corresponding conformal Noether charges, as combinations of the Hamiltonian scalar constraint, the extrinsic curvature and the 3D volume, which form a closed $\sl(2,\R)$ Lie algebra. On the other hand, this approach allows to write FLRW cosmology in terms of a AdS${}_{2}$ phase space and  a Schwarzian action. Preserving this conformal structure at the quantum level fixes the ordering ambiguities in the Wheeler-de Witt quantization and allows to formulate FLRW quantum cosmology as a CFT${}_{1}$. We show that the CFT two-points correlator is realized as the overlap of the evolution in proper time of cosmological coherent wave-packets. In particular, the two-points function is built from a vacuum state which, although not conformally invariant, coincides with the cosmological vacuum annihilated by the scalar constraint. These results suggest new perspectives in classical and quantum cosmology, among which the possibility to apply the conformal bootstrap program to quantize cosmological backgrounds.
}

\begin{document} 
\maketitle
\flushbottom

\section{Introduction}

It is well known that cosmological and black holes spacetimes share many physical and mathematical features. Classically,  information escaping beyond the cosmological horizon or falling inside the black hole  cannot be recovered. Moreover, both cosmological and black hole horizons radiate at a temperature given by their surface gravity. As a consequence, cosmological spacetimes can be considered as thermodynamical objects equipped with an entropy and a temperature, just like black holes  \cite{Bekenstein:1973ur, Bardeen:1973gs, Hawking:1974rv, Jacobson:1995ab, Gibbons:1977mu, Frolov:2002va, Davis:2003ye, Cai:2005ra, Akbar:2006kj, Cai:2008gw}. This striking feature of cosmological and black holes spacetimes suggests that their classical  mechanics as described by general relativity is only a statistical description of micro-states associated to more fundamental degrees of freedom of the underlying quantum geometry. Understanding the nature of these micro-states, and therefore the entropy of cosmological and black holes geometries, remains one major challenge of modern theoretical physics and an ultimate goal of a quantum theory of gravity.

Over the last decades, important efforts have been devoted to understanding the state counting of black holes. One major outcome was the discovery that some unexpected conformal symmetries emerge in the near-horizon region \cite{Carlip:1998wz, Carlip:1999cy, Carlip:2002be}. These conformal structures allow to import conformal field theory (CFT) results and build a universal effective conformal description of the thermodynamical properties of black holes, without relying on the details of specific quantum gravity proposals \cite{Carlip:2011vr, Carlip:2012ff} (see also \cite{Bardeen:1999px, Kunduri:2007vf, Kunduri:2013ana, Hajian:2013lna, Compere:2015mza} for the  near-horizon of extremal black holes). Moreover, additional conformal symmetries were also found for test fields propagating on the near-horizon region in various black holes spacetimes and were used to re-derive many features of these black holes from a CFT perspective, such as their spectroscopic properties, the dynamics of super-radiant modes on the Kerr background, the low energy quasi-normal modes spectrum, the emission of gravitational waves and finally their state counting\footnote{This set of results led to the Kerr/CFT conjecture and its black hole/CFT cousins \cite{Guica:2008mu}, which aim at building a consistent holographic description of black holes (see \cite{Bredberg:2011hp, Compere:2012jk} for reviews).}  \cite{Maldacena:1997ih, Bredberg:2009pv, Birmingham:2001pj, Chen:2010sn, Chen:2010ik, Castro:2010fd, Bertini:2011ga, Lowe:2011aa, Porfyriadis:2014fja, Pathak:2016vfc}. This body of results obtained so far suggests that the local physics near black hole horizons is encoded in the underlying conformal structure of the near horizon gravitational field. From this perspective, the common features of black holes and cosmological spacetimes rise a natural question concerning the latter: 
\begin{itemize}
\item Do hidden conformal symmetries exist in cosmological backgrounds ? 
\item If yes, can the thermodynamical features of cosmological spacetimes  be reproduced from an effective conformal description based on CFT techniques ? 
\end{itemize}

It is well known that similar conformal symmetries exist in the de Sitter geometry. As shown in \cite{Anninos:2011af}, test probes enjoy indeed a $\SL(2,\mathbb{R})$ structure. Quasi-normal modes and propagators of these test fields can then be obtained from the $\SL(2,\mathbb{R})$ representations\footnote{One could also mention the on-going efforts to understand the thermodynamical properties of the de Sitter space from an holographic point of view, leading eventually to the well known de Sitter/CFT conjecture \cite{Strominger:2001pn, Dyson:2002nt, Anninos:2011ui, Compere:2014cna, Neiman:2017zdr}. Yet, most of the results obtained so far concern the three-dimensional version of de Sitter. More recently, asymptotic symmetries associated to cosmological horizons have been investigated in \cite{Donnay:2019zif, Grumiller:2019fmp}.}.
 Moreover, conformal symmetries of test probes propagating on more general FRW backgrounds have also been investigated in \cite{Kehagias:2013xga}. The presence of such conformal symmetry provides a powerful tool to bootstrap the correlation functions of test probes on inflationary backgrounds, motivating the development of the cosmological bootstrap program to cosmological perturbations. See \cite{Kehagias:2012pd,  Kehagias:2015jha, Kehagias:2017rpe} and \cite{Arkani-Hamed:2018kmz, Baumann:2019oyu} for different investigations on this point. Up to now, these investigations have focused on the quantization of test fields propagating on classical cosmological backgrounds. A natural question is whether a more fundamental conformal structure exists at the level of the background spacetime itself, allowing the bootstrap its quantization. 

From the point of view of quantum gravity, the quantization of cosmological backgrounds represent the simplest models one can build. By freezing all the gravitational degrees of freedom but the scale factor, and considering a non-trivial matter content such as a scalar field, one obtains a simple symmetry-reduced model which allows to test different quantization schemes. Performing a quantization using the canonical approach or the path integral one, one obtains a simple quantum theory encoding the quantum properties of a homogeneous region of space. Despite the apparent simplicity of such quantum cosmology, it is worth pointing that one still faces the crucial challenges inherent to the quantization of the gravitational field, such as the timeless nature of the Wheeler-De Witt equation and the realization of the quantum covariance. Most efforts in quantum cosmology have been devoted to provide suitable initial conditions for the quantum universe \cite{Hartle:1983ai, Vilenkin:1982de, Feldbrugge:2017kzv, Vilenkin:2018dch}, as well as to discuss the fate of the classical Big Bang singularity at the quantum level (see \cite{Bojowald:2015iga, Bojowald:2012xy, Bojowald:2010cj, Halliwell:2002cg, Vilenkin:1994ua, Hartle:1997hw} for reviews). In this work, we would like to point that, if the thermodynamics and quantum properties of cosmological spacetimes can indeed be reformulated in CFT terms, quantum cosmology certainly provides the simplest framework to study these questions. 

As a matter of fact, it turns out that recent investigations in canonical quantum cosmology have revealed an unexpected conformal structure in the phase space of the homogeneous and isotropic Einstein-Scalar system\footnote{Related works in quantum cosmology based on a $\sl(2,\mathbb{R})\sim \su(1,1)$ hidden structure in loop quantum cosmology can be found in \cite{Bodendorfer:2019wik, Bodendorfer:2018csn, Livine:2012mh, Bojowald:2007bg}.} \cite{BenAchour:2019ywl, BenAchour:2018jwq, BenAchour:2017qpb}. This structure is very similar to the one appearing in the simplest example of a one-dimensional $\text{CFT}$, the well-known conformal mechanics introduced by de Alfaro, Fubini and Furland  (dAFF) in the mid seventies \cite{deAlfaro:1976vlx}. See \cite{Andrzejewski:2011ya, Andrzejewski:2015jya, Okazaki:2017lpn, Cadoni:2000iz, Khodaee:2017tbk, Carinena:2017zfy} for more details on this system. This simple mechanical model have found applications in a wide range of physical systems, see \cite{Mignemi:2001uz, Camblong:2003mz, Clement:2001ny, Gaiotto:2004ij, Camblong:2004ye} for examples, and has recently regained attention due to its potential role in the the $\text{AdS}_2/\text{CFT}_1$ conjecture \cite{Strominger:2003tm, Spradlin:1999bn, Azeyanagi:2007bj, Hartman:2008dq, Chamon:2011xk, Axenides:2013iwa, Pinzul:2017wch, Mezei:2017kmw,Gupta:2017xex, Grumiller:2017qao, Kolekar:2018sba, Sarosi:2017ykf}. The goal of the present work is to show that the homogeneous and isotropic Einstein-Scalar system enjoys indeed the very same conformal symmetry under the 1d conformal group $\SL(2,\mathbb{R})$ than the well known dAFF model \cite{deAlfaro:1976vlx}. As such, conformal mechanics also plays a crucial role for classical and quantum cosmology. In the following, we shall therefore investigate deeper the conformal structure found in \cite{BenAchour:2019ywl, BenAchour:2018jwq, BenAchour:2017qpb} for FLRW cosmology and clarify the concrete mapping with the dAFF model.

More precisely, we  focus our attention to the homogeneous and isotropic Einstein-Scalar cosmological system. Surprisingly, we shall find that, additionally to the standard invariance under time reparametrization, the action of this system turns out to also be invariant under Mobius transformation of the proper time  provided the scale factor transforms as a primary field. As a consequence, the cosmological action can be recast in the form of a Schwarzian action for cosmology. Moreover, following Noether's theorem, the existence of this new symmetry implies the existence of three new conserved charges in FLRW cosmology, associated to the translation, dilatation and special conformal transformations of the time coordinate. From the gravitational point of view, these new conserved charges have two surprising properties: they are time-dependent and they correspond to bulk currents. At the Hamiltonian level, we identify these Noether charges as (the initial conditions for) the extrinsic curvature $\cC$, the 3D volume $v$ and the scalar Hamiltonian constraint $\cH$, and show that  they form an $\sl(2,\mathbb{R})$ Lie algebra.
This clarifies the physical meaning of the so-called CVH algebra for FLRW cosmology introduced earlier in \cite{BenAchour:2019ywl, BenAchour:2018jwq, BenAchour:2017qpb} and provides a more complete picture of this structure.

This unexpected structure of the homogeneous and isotropic Einstein-Scalar system allows to build a straightforward mapping with the $\SL(2,\mathbb{R})$-invariant conformal mechanic of de Alfaro, Fubini and Furlan (dAFF) \cite{deAlfaro:1976vlx}. Indeed, a simple reformulation of the gravitational system shows that the dAFF mechanics corresponds to the cosmological action written in proper time.
Additionally, we show that the cosmological phase space is naturally endowed with the $\text{AdS}_2$ metric, which allows to identify  the physical trajectories on the cosmological phase space as $\text{AdS}_2$ geodesics. 

At the classical level, these findings show that, as anticipated in \cite{BenAchour:2019ywl, BenAchour:2018jwq}, the homogeneous and isotropic symmetry reduced Einstein-Scalar system is a  covariant $\text{CFT}_1$. This is rather surprising as GR is well known to not be conformally invariant. However, it appears that for a specific choice of foliation, this gravitational system enjoys indeed an exact $\SL(2,\mathbb{R})$ symmetry associated to bulk conserved currents. This provides a very simple example of an exact conformal symmetry in a symmetry-reduced sector of 4d General Relativity.

At the quantum level, this newly found conformal structure turns out to have far-reaching consequences, as expected. Preserving the $\sl(2,\mathbb{R})$ charge algebra in a  canonical quantization scheme, \`a la Wheeler-de Witt,  provides a powerful criterion to narrow the quantization ambiguities and determine the operator ordering (see \cite{BenAchour:2019ywl, BenAchour:2018jwq, BenAchour:2017qpb} about  preserving the $\sl(2,\mathbb{R})$ structure for alternative quantization schemes).
In the quantum theory, the natural question is then how do the usual CFT correlation functions appear in the quantum cosmology framework.
We will show that the overlap between  coherent wave packets evolving in time actually reproduces the universal form of the two-point CFT correlator. In order to derive this result, we will follow closely the approach proposed in \cite{Chamon:2011xk} for conformal quantum mechanics. A crucial point is that the CFT correlator cannot be recovered in the standard way because there are no conformally invariant vacuum state in the present one-dimensional setup. However, as shown for conformal quantum mechanics in \cite{Chamon:2011xk}, it is still possible to identify non-normalizable states together with a quasi-primary operator such that both conspire to lead to the desired form for the correlator. The same technics will be used here. From a more general perspective, the reformulation of this quantum cosmology in term of a $\text{CFT}_1$ suggests that one could solve the model based solely on symmetry arguments and import the conformal bootstrap program \cite{Poland:2018epd}  within quantum cosmology.

\medskip

This paper is organized as follows. In section-\ref{sec2.1}, we present the Mobius invariance of the homogeneous and isotropic symmetry reduced action of the Einstein-Scalar system and write the cosmological action in term of the Schwarzian derivative in section-\ref{sec2.2}. In section-\ref{sec2.3}, we derive the associated Noether charges. In section-\ref{sec2.4}, we establish the mapping with the dAFF model. In section-\ref{sec3}, we investigate the Hamiltonian realization of the conformal symmetry and show its relation to the  CVH algebra introduced in \cite{BenAchour:2017qpb, BenAchour:2019ywl, BenAchour:2018jwq}. We then show that the Noether charges form a $\sl(2,\R)$ Lie algebra. Section-\ref{evolution} is dedicated to showing that the cosmological dynamics can be fully integrated using the $\SL(2,\mathbb{R})$ group structure while Section-\ref{ads2} derives the cosmological trajectories from the $\text{AdS}_2$ geometry. Finally section-\ref{sec:quantization} tackles  the quantum theory. In section-\ref{sec:wdw}, we present the  Wheeler-De Witt quantization preserving the $\sl(2,\mathbb{R})$ structure of the charge algebra. Finally,  section-\ref{sec:CFT} presents the derivation of the CFT 2-point function in the quantum cosmology seting. We conclude by a discussion on the research directions opened by our results on the conformal invariance of FLRW cosmology.

\section{$\SL(2,\R)$ Symmetry of FLRW cosmology} 

\label{sec1}

In this section, we review the lagrangian and hamiltonian formulation of FRLW cosmology filled with a massless scalar field, setting our notation for the rest of this work. Then, we show that this simple cosmological system, on top of the invariance under time reparametrization, enjoys an additional invariance under the 1d conformal group, the scale factor transforming as a quasi-primary field under Mobius transformation of the time coordinates. The covariance under Mobius transformation of the cosmological action is shown to leads to three conserved Noether's charges. At the hamiltonian level, we show that these charges form an $\sl(2,\mathbb{R})$ structure on the phase space, and the cosmological dynamics can be fully reproduced by integrating the flow generated by these three Noether's charges. As a result, this simple cosmological system appears as a covariant version of the de Alfaro-Fubini-Furland conformal mechanics developed in the seventies \cite{deAlfaro:1976vlx}. This unexpected mapping between the simplest $\text{CFT}_1$ model and the homogeneous and isotropic sector of General Relativity allows to recast classical cosmology as a simple one dimensional conformal field theory and pave the way for a straightfroward quantization in term of $\SL(2,\mathbb{R})$ representations. In Appendix-\ref{app:Lambda}, we underline that this conformal structure extends beyond this simple system, including the $\text{(A)dS}$ cosmology and the Bianchi I universe.

\subsection{Lagrangian formulation: Symmetries and Noether Charges}

Let us consider the FRLW geometry in the isotropic and homogeneous flat slicing
\be
\label{le}
\rd s^2 = - N^2(t) \rd t^2 + a^2(t) \delta_{ij} \rd x^i \rd x^j 
\ee
For an homogeneous minimally coupled massless scalar field $\phi$, the symmetry reduced action of the Einstein-Scalar system integrated over a cell of volume $V_{\circ}$ reads:
\be
\label{actionFRLW}
\cS[N, a, \dot{a}, \dot{\phi}]
 =
V_{\circ} \int_{\mathbb{R}} \rd t
 \,\left[
 a^3\frac{\dt{\phi}^2}{2N} 
- \frac{3}{8\pi G}\frac{a \dt{a}^2}{N} 
 \right]
 \,,
\ee
where the primes stand for the time derivative, $\dt{a}=\rd a /\rd t$ and $\dt{\phi}=\rd \phi /\rd t$.
Here we do not discuss the (Gibbons-Hawking) boundary terms living on the spatial boundary of the fiducial cell. They should not play any important role in the homogeneous dynamics, but should become relevant in a more general setting.

This action depends on two lengths scales. On the one hand, the scale $\ell_{\circ} = V_{\circ}^{1/3}$ encodes the size of the fiducial cell to which we restrict the spatial integration to avoid divergences. It corresponds to an infrared cut-off of the system. This scale plays a crucial role when working with such finite homogeneous region of space, setting the scale at which inhomogeneities can develop. On the other hand, the Planck length $\ell_{\text{P}} = \sqrt{8\pi G}$ signals the high energy UV regime

As it is well known, this theory
enjoys a symmetry under arbitrary reparametrization of the time coordinate $t$, which manifest through the invariance under arbitrary rescaling of the lapse function $N(t)$.
More precisely, time reparametrization corresponds to the transformations
\be
\left|
\begin{array}{lcrcl}
t&\longmapsto& \tt&=&f(t) \\
\rd t&\longmapsto& \rd \tt&=&h(t) \rd t \\
N&\longmapsto&\tN(\tt)&=& h^{-1}(t) N(t) \\
a &\longmapsto& \ta(\tt)&=&a(t)\\
\phi&\longmapsto& \tphi(\tt)&=&\phi(t)
\end{array}
\right.
\qquad\textrm{with the Jacobian}\quad
h(t)\equiv f'(t)=\f{\rd f}{\rd t}
\,.
\ee
Due to this invariance, it is convenient physically to absorb the lapse in the time coordinate and write the action and the resulting equations of motion in terms of the proper time $\tau$ defined by $\rd \tau = N \rd t$. The action then reads:
\be
\label{actionproper}
\cS[N, a, \dot{a}, \dot{\phi}]
=
V_{\circ} \int_{\mathbb{R}} \rd \tau
 \,\left[
\frac{ a^3\dot{\phi}^2}{2} 
- \frac{3\,a \dot{a}^2}{8\pi G}
 \right]
 =
 V_{\circ} \int_{\mathbb{R}} \rd \tau
 \,\left[
\frac{ a^3\dot{\phi}^2}{2} 
- \frac{1}{6\pi G}\left(\f{\rd a^{3/2}}{\rd\tau}\right)^{2}
 \right]
 \,,
\ee
where the dot notation now stands for the  derivative with respect to the proper time, $\dot{a}=\rd a /\rd \tau$ and $\dot{\phi}=\rd \phi /\rd \tau$.
Then the variations with respect to  the scale factor, the scalar field and the lapse function lead to the standard Friedman equations:
\begin{align} 
\label{eo1} 
\cE_a & = 2 a \ddot{a} + \dot{a}^2 + 4 \pi G a^2 \dot{\phi}^2 =0 \,,\\
\label{eo2} 
\cE_{\phi} & = \rd_{\tau}(a^{3}\dot{\phi})=3 a^2 \dot{a} \dot{\phi} + a^3 \ddot{\phi} = 0 \,,\\
\label{eo3} 
\cE_N & = a \dot{a}^2 - \frac{4\pi G}{3} \; a^3 \dot{\phi}^2 =0 \,.
\end{align}
%
The first one corresponds to the acceleration equation, the second to the continuity equation while the third one is the standard first Friedman equation.
Notice that the dynamics depends on the Planck length $\ell_{\text{P}} = \sqrt{8\pi G}$ but not on the scale $\ell_{\circ}$ associated to the fiducial cell. 

\subsubsection{Mobius covariance of the action} 

\label{sec2.1}

On top of the invariance under 1D diffeomorphism given by time reparametrization, it turns out that the cosmological action admits an unexpected time-dependent conformal symmetry.
This invariance is clearer written in proper time. Starting from the action \eqref{actionproper} given above, we perform a Mobius transformation of the proper time:
\begin{align}
\label{mobius}
\tau&\,\longmapsto\, \ttau=\frac{\alpha \tau + \beta }{\gamma\tau + \delta}
\qquad \text{with}\quad
\alpha\delta-\beta\gamma =1
\,, \\
a &\,\longmapsto\, \ta(\ttau)=\f{a(\tau)}{(\gamma\tau +\delta)^{{2/3}}} \,,
\nn\\
\phi&\,\longmapsto\, \tphi(\ttau)=\phi(\tau)
\,.\nn
\end{align}
In particular, this implies the following Jacobian and transformation of the time derivatives:
\begin{align}
 \rd\ttau & = \frac{ \rd\tau}{(\gamma \tau + \delta)^2}
 \,, \\
 \frac{\rd \, \tphi}{ \rd\ttau} & =  ( \gamma \tau + \delta)^{2} \frac{\rd \, \phi}{ \rd\tau}
 \,,\nn \\
\left( \frac{\rd \, \ta^{3/2}}{ \rd\ttau}\right)^2
& =
( \gamma \tau + \delta)^{2}
\left[
\left( \frac{\rd \,a^{3/2}}{\rd \tau}\right)^2 - \gamma \frac{\rd}{\rd \tau} \left( \frac{a^3}{\gamma \tau + \delta}\right)
\right]
\,.\nn
\end{align}
The latter is the key identity to prove the invariance of the action:
\begin{align}
\int \rd\ttau \cL \left[ \ta(\ttau),\tilde{\dot{a}}(\ttau), \tilde{\dot{\phi}}(\ttau) \right]
&= 
V_{\circ} \int \rd \ttau
 \,\left[
\frac{ \ta^3}{2} \left(\f{\rd \tphi}{\rd\ttau}\right)^{2}
- \frac{1}{6\pi G}\left(\f{\rd \ta^{3/2}}{\rd\ttau}\right)^{2}
 \right]
\nn\\
&=
\int \rd\tau\,\left(  \cL \Big[ a(\tau),\dot{a}(\tau), {\dot\phi}(\tau) \Big]+\f{\rd F}{\rd \tau}\right)
\end{align} 
with the total derivative term is given by:
\be
F(\tau)=\f{V_{\circ}}{6\pi G}\f{\gamma}{(\gamma \tau +\delta)}\,a^{3}
\,.
\ee
Therefore, while the Lagrangian is not invariant, the action is modified by a total derivative and the equation of motion are therefore unaffected by the above  transformations. This shows that the homogeneous and isotropic Einstein-Scalar action enjoys an additional Mobius covariance on top of the time reparametrization freedom of the action.

These transformations  can be understood as conformal transformations on the metric:
\be
\label{transfo}
\rd s^2 = - \rd\tau^{2}+ a^2(\tau) \delta_{ij} \rd x^i \rd x^j 
\quad\longmapsto\quad
\widetilde{\rd s}^{2}
=
 - \f{\rd\tau^{2}}{(\gamma \tau +\delta)^{2}}
 +  \f{a^2(\tau)\delta_{ij} \rd x^i \rd x^j }{(\gamma \tau +\delta)^{4/3}} 
\ee
It is worth pointing that such new continuous symmetry for homogeneous cosmology is a rather unexpected property since general relativity is not conformally invariant. This goal of this work is to discussed in detail this new conformal structure of the cosmological action.

\medskip

It is interesting to translate this Mobius transformation in proper time in terms of coordinate time and lapse. Indeed, since the $\rd\tau=N\rd t$, the relation between $\tau$ and $\ttau$ can be achieved through different mappings from $(t,N)$ to $(\tt, \tN)$. All these implementations in proper time are related to each other by time reparametrizations. The simplest realization is assuming that the coordinate time does not change, $\tt=t$, and absorb all the Mobius transformation in the transformation of the lapse function:
\begin{align}
\label{mobiusint}
t&\,\longmapsto\, \tt=t
\,,\nn\\
a &\,\longmapsto\, \ta(t)=\f{a(t)}{(\gamma\tau +\delta)^{{2/3}}} \,,\nn\\
\phi&\,\longmapsto\, \tphi(t)=\phi(t)\,,\nn\\
N&\,\longmapsto\, \tN(t)=\f1{(\gamma\tau +\delta)^{2}}\,N(t)
\,.
\end{align}
An important remark is that these transformations\footnotemark{} explicitly involve the proper time $\tau(t)=\int \rd t\, N$ as a function of the coordinate time $t$.
\footnotetext{
The transformation laws \eqref{mobiusint} in coordinate time seem only to depend on the two parameters $\gamma$ and $\delta$, although the actual Mobius transformation is defined by three parameters, $\alpha$, $\beta$, $\gamma$ and $\delta$ being related by the constraint $\alpha\delta-\beta\gamma=1$. Nevertheless, the definition of the proper time as an integral, $\tau=\int^{t}_{t_{0}}  N$ involve a further parameter - the initial time $t_{0}$- which provides the missing parameter. This corresponds to choosing the event at which the proper time vanishes, $\tau(t_{0})=0$.}

\subsubsection{Conformal transformation and Schwarzian action for Cosmology} 
\label{sec2.2}

Before computing the Noether charges associated with the Mobius invariance, we would like to consider general conformal transformations and show how the FLRW action is related to the Schwarzian action.

Starting with the FLRW action \eqref{actionproper} as an integral in proper time, we perform an arbitrary transformation of the proper time and assume that the scale factor transforms as a primary field of weight $\f13$ (so that the 3D volume is of weight 1):
\be
\label{transfo1}
\left|
\begin{array}{lcl}
\tau&\mapsto&\ttau=f(\tau)
\,,\vspace*{1mm}\\
\rd\tau&\mapsto&\rd\ttau=\dot{f}\rd\tau\equiv h(\tau)\rd\tau
\,,\vspace*{1mm}\\
a&\mapsto&\ta(\ttau)=h^{\sfrac{1}{3}}(\tau)a(\tau)
\,,
\end{array}
\right.
\quad
\frac{\rd\ta^{\sfrac{3}{2}}}{\rd\ttau}
=
\f1h\f{\rd (h^{\sfrac{1}{2}}a^{\sfrac{3}{2}})}{\rd\tau}
=
h^{-\sfrac{1}{2}}\frac{\rd a^{\sfrac{3}{2}}}{\rd\tau}
+\f12h^{-\sfrac{3}{2}}\dot ha^{\sfrac{3}{2}}
\,.
\ee
A straightforward calculation allows to show that the variation of the action is given exactly by a Schwarzian action up to a total derivative:
\be
\label{transfoaction}
\cS[\ta, \tilde{\dot{a}}, \tilde{\dot{\phi}}] - \cS[a, \dot{a}, \dot{\phi}]
=
\frac{V_{\circ}}{12\pi G}
\int \rd\tau
\left[
Sch[f] a^3- \frac{d}{d\tau} \left( \frac{\ddot{f}}{\dot{f}} a^3\right)
\right]
\,,
\ee
where $Sch[\cdot]$ denotes the Schwarzian derivative, defined as:
\be
Sch[f] = \frac{ \dddot{f}}{\dot{f}} - \frac{3}{2} \left( \frac{\ddot{f}}{\dot{f}}\right)^2
=
 \frac{ \ddot{h}}{{h}} - \frac{3}{2} \left( \frac{\dot{h}}{{h}}\right)^2
 \,.
\ee
This means that these transformations are a symmetry of the cosmological action if and only if the Schwarzian derivative vanishes, $Sch[f] =0$, which is true only for Mobius transformations:
\be
\label{transfo2}
f(\tau) = \frac{\alpha \tau + \beta }{\gamma\tau + \delta}
\qquad \text{with}\quad
\alpha\delta-\beta\gamma =1
\,.
\ee
This confirms that FLRW cosmology is invariant under $\sl(2,\mathbb{R})$ conformal transformations, on top of the 1D diffeomorphisms defined by time reparametrizations.  

\subsubsection{The associated Noether's charges}
\label{sec:noethervariation}
\label{sec2.3}

Thanks to the Noether's theorem, we known that any continuous symmetry of the action is associated to the existence of conserved charges. Let us now derive these charges.
The Mobius transformations of the proper time  (\ref{mobius}) can be decomposed into combinations of translations, dilatations and special conformal transformations, whose infinitesimal actions read:
\be
\begin{array}{lcl}
\ttau  = \tau + \epsilon &\qquad& \text{translation ($\alpha=\delta=1,\gamma=0, \beta=\eps$)}\,,
\vspace*{1mm}\\
\ttau  = \tau + \epsilon \tau &\qquad&\text{dilatation ($\beta=\gamma=0,\alpha=\delta^{-1}=\sqrt{1+\eps}$)}\,,
\vspace*{1mm}\\
\ttau  = \tau + \epsilon \tau^2 &\qquad &\text{special conformal transformation  ($\beta=0,\alpha= \delta=1,\gamma=-\eps$)}\,.
\end{array}
\ee
We shall therefore derive the Noether's charges associated to each of these elementary transformations.
\begin{itemize}
\item \textit{Translation:} \\
Under an infinitesimal translation, the proper time  $\tau$ is simply displaced with the offset, while the fields $a(\tau)$ and $\phi(\tau)$ are simply reparametrized. This reproduces a straightforward infinitesimal time reparametrization:
\be
\left|
\begin{array}{lcrcl}
\tau&\longmapsto& \ttau&=&\tau+\eps \\
a &\longmapsto& \ta(\ttau)&=&a(\tau)\\
\phi&\longmapsto& \tphi(\ttau)&=&\phi(\tau)
\end{array}
\right.
\ee
The infinitesimal field variations are:
\begin{align}
\label{itrans}
\delta_\eps^{t} a & = \ta(\tau) - a(\tau) = - \epsilon \dot{a}\,, \\
\delta_\eps^{t} \phi & = \tphi(\tau) - \phi(\tau) = - \epsilon \dot{\phi}
\,.
\end{align}
Computing the variation of the action at leading order in $\eps$ gives us the corresponding Noether charge:
\be
\delta_{\eps}S=-\eps V_{\circ}\int \rd\tau\,
\f{\rd}{\rd\tau}\left(
\f12 a^{3}\dot{\phi}^{2}-\f3{8\pi G}a\dot{a}^{2}
\right)
\,.
\ee
The associated Noether charge is thus given by
\be
Q_{-} = V_{\circ} \left[\frac{1}{2} a^3 \dot{\phi}^2 - \frac{3}{8\pi G} a \dot{a}^2 \right]\,,
\ee
which  turns out to be exactly the Hamiltonian as we will see in the next section. We check that its time derivative can be entirely expressed in terms of the equations of motion  (\ref{eo1}) and (\ref{eo2}) and thus vanishes on-shell:
\be
\dot{Q}_{-}
=
V_{\circ}\,
\left[
\dot{\phi}\cE_{\phi}-\frac{3}{8\pi G} \dot{a} \; \cE_a \right]
\hat{=}\,0
\,,
\ee
where the equality $\hat{=}$ assumes the equations of motion $\cE_{a}=\cE_{\phi}=\cE_{N}=0$.

\item \textit{Dilatation:}\\
Consider now the dilatation by a factor $\lambda = 1 + \epsilon$ with $\epsilon \ll 1$ which transforms the proper time and the fields as
\be
\left|
\begin{array}{lcrcl}
\tau&\longmapsto& \ttau&=&(1+\eps)\tau \\
a &\longmapsto& \ta(\ttau)&=&(1+\eps)^{\sfrac{1}3}a(\tau)\\
\phi&\longmapsto& \tphi(\ttau)&=&\phi(\tau)
\end{array}
\right.
\ee
Under such infinitesimal transformations, the field variations are;
\begin{align}
\label{idil}
\delta_\eps^{d} a & = \ta(\tau) - a(\tau) = \epsilon \left( \frac{a}{3} - \tau \dot{a}\right)  \\
\delta_\eps^{d} \phi & = \tphi(\tau) - \phi(\tau) = - \epsilon  \tau \dot{\phi}
\end{align}
which leads to the following Noether charge:
\be
Q_{0} = V_{\circ}\left[
\tau\left(
\frac{1}{2} a^3 \dot{\phi}^2
- \frac{3}{8\pi G} a \dot{a}^2
\right)
 + \frac{1}{4\pi G} a^2 \dot{a}
\right]
=\tau Q_{-} + \frac{V_{\circ}}{4\pi G} a^2 \dot{a}
\ee
The conservation of this charge can be checked by a direct calculation:
\be
\frac{\rd Q_{0}}{\rd\tau}
=
V_{\circ}\left[
\tau\dot\phi\eps_{\phi}
+\f3{8\pi G}\left(\f a3-\dot a\tau\right)\eps_{a}
\right]
\hat{=}\,0
\,.
\ee
\item \textit{Special conformal transformations:}\\
Consider now the special conformal transformations which consist in an inversion $\tau\mapsto -1/\tau$, followed by a translation and finally a second inversion. Under such infinitesimal special conformal  transformations, one has
\be
\left|
\begin{array}{lcrcl}
\tau&\longmapsto& \ttau&=& \frac{\tau}{1+ \epsilon \tau} \simeq \tau - \epsilon \tau^2 + \cO(\epsilon^2) \,, \vspace*{1mm}\\
a &\longmapsto& \ta(\ttau)&=&(1+\epsilon \tau)^{-2/3}\,a(\tau)\sim\left(1-\f23\eps\tau\right)a(\tau) \,, \vspace*{1mm} \\
\phi&\longmapsto& \tphi(\ttau)&=&\phi(\tau)\,.
\end{array}
\right.
\ee
One obtains the following infinitesimal field variations:
\begin{align}
\label{ist}
\delta_\eps^{s} a &= \ta(\tau) - a(\tau) \sim \epsilon \tau \left(  \tau\dot{a} - \frac{2}{3} a\right) \,,\\
\delta_\eps^{s} \phi &= \tphi(\tau) - \phi(\tau)  \sim  \epsilon \tau^{2} \dot{\phi}\,,
\end{align}
from which one can derive  the associated Noether charge:
\beq
\label{NC-ST}
Q_{+} &=&
V_{\circ}\,\left[
-\tau^2 \left( \frac{1}{2} a^3 \dot{\phi}^2- \frac{3}{8\pi G} a \dot{a}^2 \right)
- \tau \frac{1}{2\pi G} a^2 \dot{a}
+ \frac{1}{6\pi G} a^3
\right] \nn\\
&=&
\tau^{2}Q_{-}-2\tau Q_{0}+\f{V_{\circ}}{6\pi G}a^{3}
\,.
\eeq
The conservation of this charge can be checked directly assuming the equations of motion:
\be
\frac{\rd Q_{+}}{\rd\tau}
=
\tau^{2}\dot Q_{-}-2\tau \dot Q_{0}
\,\hat=\,0
\,.
\ee
\end{itemize}
As expected, the new conformal symmetry under Mobius transformations of the action (\ref{actionFRLW}) leads to three time-dependent charges which are conserved in time. Looking carefully at their structure, it appears that the charge $Q_{-}$ associated to the translations should correspond the Hamiltonian, while $Q_{0}$ and $Q_{+}$ seems to be the evolving constants of motion respectively for $a^{2}\dot a$ and $a^{3}$.
In the next section dedicated to the Hamiltonian analysis of the system, we will confirm this interpretation and further show that these three Noether charges indeed form an $\sl(2,\mathbb{R})$ algebra as expected. This will turn out to have far reaching consequences both classically and quantum mechanically.

Nevertheless before proceeding to the canonical analysis of the cosmological action,we would like to show how this conformal structure becomes apparent through a simple change of variables. This will provides a direct mapping between FLRW cosmology and to the  conformal mechanics developed by de Alfaro, Fubini and Furland \cite{deAlfaro:1976vlx}, which is now involved in the study of the AdS${}_{2}$/CFT${}_{1}$ correspondence. 

\subsubsection{Mapping to conformal mechanics}
\label{DFF}
\label{sec2.4}

The close analogy of the above conformal symmetry of the action with the well known conformal mechanics introduced by de Alfaro, Fubini and Furland \cite{deAlfaro:1976vlx} suggests that one could map the two systems. Indeed, it turns out that this analogy can be made explicit by a change of variables. Let us therefore introduce the new set of variables
\be
q(\tau) = \sqrt{\frac{V_{\circ} }{6\pi G}}\,a^{\sfrac 32}(\tau), \;  \qquad \beta =-  \frac{1}{12 \pi G} \left( \frac{a^3 \dot{\phi}}{2}\right)^2
\ee
with physical dimension $[q] = L^{1/2}$ and $[\beta] = 1$. From the continuity equation (\ref{eo2}), we know that $\beta$ is a constant of motion. In term of these variables,
the FLRW cosmological action reduces\footnotemark{} to:
\begin{align}
\label{actiondAFF}
\cS_{\beta}[q, \dot{q}] = \int_{\mathbb{R}} 
\rd\tau\,\left[ \dot{q}^2 - \frac{\beta}{q^2}  \right]
\
\end{align}
\footnotetext{
Effectively assuming that $\beta$ is a constant amounts to solving the equations of motion for the scalar field $\phi$ and inserting the on-shell solution back in the action to obtain an action for the sole scale factor $a$.}
This recovers the exact action of conformal mechanics introduced by dAFF in \cite{deAlfaro:1976vlx}. From this point of view, the analogy with the dAFF model is made explicit. Now, it is well known that this system enjoys an invariance under the 1d conformal group, and provides an example of a 1D conformal field theory \cite{deAlfaro:1976vlx} It has been studied intensively over the last decades, see for examples \cite{Andrzejewski:2011ya, Andrzejewski:2015jya, Okazaki:2017lpn, Cadoni:2000iz, Khodaee:2017tbk, Carinena:2017zfy} as well as \cite{Mignemi:2001uz, Camblong:2003mz, Clement:2001ny, Gaiotto:2004ij, Camblong:2004ye} for several applications to black holes. Its equation of motion reads simply
\be
\cE_q = \ddot{q} - \frac{2\beta}{q^3} =0
\ee
The Noether charges associated to the conformal invariance are given by \cite{deAlfaro:1976vlx}:
\begin{align}
Q_{-} & = \frac{\dot{q}^2}{2} + \frac{\beta}{q^2}
\,,\\
Q_{0} & = \frac{1}{2} q \dot{q} - \tau \left( \frac{\dot{q}^2}{2} + \frac{\beta}{q^2}\right)
= \frac{1}{2} q \dot{q} - \tau Q_{-}
\,,\\ 
Q_{+} & = \frac{q^2}{2} - \tau q \dot{q} + \tau^2 \left(\frac{\dot{q}^2}{2} + \frac{\beta}{q^2}\right)
=\frac{q^2}{2} - 2\tau Q_{0}-\tau^{2}Q_{-}
\,,
\end{align}
such that their time derivatives are all proportional to the equation of motion and thus vanish on-shell:
\be
\frac{\rd Q_{-}}{\rd \tau}  = {\dot{q}} \cE_q
\,,\qquad
\frac{\rd Q_0}{\rd \tau}  = \frac{1}{2}\left( q-2\tau\dot{q}  \right) \cE_q
\,,\qquad
\frac{\rd Q_{+}}{\rd \tau}  = \left( \tau^2 \dot{q}-\tau q \right) \cE_q
\,.
\ee
Consequently, the cosmological action (\ref{actionFRLW}) can be mapped to the dAFF model. We point out that it is nevertheless quite surprising that the homogeneous and isotropic sector of the Einstein-Scalar system reduces to such one-dimensional conformal field theory. 

Before concluding this section, let us discuss a subtlety of the conformal mechanics. In two dimensions, a CFT is associated to a Virasoro symmetry and thus an infinite set of conserved charges generating the 2d diffeomorphism. It was shown in \cite{Kumar:1999fx, Cacciatori:1999rp, Mignemi:2000cv} that conformal mechanics is also equipped with an infinite set of conserved charges which form an $w_{\infty}$ algebra, a subsector of which is the Virasoro algebra. However, because the system contains only a finite number of degrees of freedom, contrary to the two dimensional case, the charges forming this infinite set are not independent and boil down to combination of the three conserved charges generating the $\SL(2,\mathbb{R})$ algebra. This is a special property to keep in mind when dealing whith such one dimensional conformally invariant system. It would be interesting to understand how this infinite set of charges can be constructed in our cosmological model. We leave this for future work.

 At the quantum level, another subtlety of this system is that contrary to 2d CFT, there is no vacuum state annihilated by the three $\SL(2,\mathbb{R})$ generators. However, as shown in \cite{Chamon:2011xk}, it is still possible to find a suitable operator such that its two points function admits the standard form of the CFT two point correlator. The construction of the three and four points functions have also been investigated in \cite{Jackiw:2012ur}. In the second part of this paper, devoted to the quantum theory, we will show that the desired two points function can be constructed in quantum cosmology, and the vacuum, while not  $\SL(2,\mathbb{R})$-invariant, coincides with the cosmological vacuum annihilated by the scalar constraint, finding therefore a natural interpretation in this cosmological context.

Having presented the conformal invariance of the cosmological system (\ref{actionFRLW}) at the Lagrangian level, we turn now to the Hamiltonian formulation, which will allows us to compute the charge algebra and exhibit the underlying $\sl(2,\mathbb{R})$ structure of the phase that we shall use as a guiding line through the quantization.

\subsection{Hamiltonian formulation} 
\label{sec3}
Let us now proceed to the canonical analysis of the cosmological action \eqref{actionFRLW}:
\be
\cS[N, a, a', {\phi'}] 
=
V_{\circ} \int_{\mathbb{R}} \rd t
 \,\left[
 a^3\frac{\dt{\phi}^2}{2N} 
- \frac{3}{8\pi G}\frac{a \dt{a}^2}{N} 
 \right]
\,,
\nn
\ee
where $a'$ and $\phi'$ denote the derivative with respect to the coordinate time $t$.
Its Hamiltonian form is obtained by a Legendre transform.
We compute the conjugate momenta to the configuration variables $\left(a, \phi\right)$:
\be
\pi_a =\f{\pp \cL}{\pp a' }
= -\frac{3V_{\circ}}{4\pi G} \frac{a {a'}}{N}
\,, \qquad
\pi_{\phi}
=\f{\pp \cL}{\pp \phi' }
= \frac{a^3 V_{\circ}}{N} {\phi'}
\,,
\ee
with physical dimensions $[\pi_a] = [a] = 1$ and $[\pi_{\phi}] = L$.
The symplectic structure is given by the canonical Poisson brackets between pairs of conjugated variables:
\be
\{ a, \pi_a\} = 1\,, \qquad \{ \phi, \pi_{\phi}\} = 1 
\ee
The cosmological action can then be written as:
\be
S[N, a, \pi_a, \phi, \pi_{\phi}]
=
\int \rd t\,
 \left(
 \pi_a a' + \pi_{\phi} {\phi'} - H
 \right)
\ee
with the Hamiltonian given by:
\be
 H=N\cH\,,\qquad
 \cH
 =
 \f 1{V_{\circ}}\left[
 \frac{\pi^2_{\phi}}{2 a^3} - \frac{2\pi G}{3} \frac{\pi^2_a}{a}
 \right]
 \,.
\ee
As usual, the lapse function $N$ is  not a dynamical variable and plays the role of a Lagrange multiplier enforcing the scalar constraint $\cH \simeq 0$, with physical dimension  $[\cH] =L^{-1}$.
While the Hamiltonian generates the evolution in the coordinate time $t$, the scalar constraint $\cH$ generates the evolution in proper time $\tau$. So, for an arbitrary observable $\cO$ of the cosmological phase space, we have:
\be
\cO'=\f{\rd \cO}{\rd t}=\{\cO,H\}
\,,\qquad
\dot\cO=\f{\rd \cO}{\rd \tau}=\f1N\cO'=\{\cO,\cH\}
\,.
\ee

It is convenient to introduce a new set of canonical variables. We define the rescaled 3d volume as:
\be
v = \int d^3x \sqrt{\gamma} = a^3 V_{\circ}
\ee
such that $[v] = L^3$. Its associated canonical variable is given by the Hubble factor $b$ which reads 
\be
b = - \frac{1}{3V_{\circ}} \frac{\pi_a}{a^2} = \frac{1}{4\pi G} \frac{{a'}}{Na}\,, \qquad \text{such that} \quad \{ b, v\} = 1
\,.
\ee
Its physical dimension is as expected an inverse volume, $[b] = L^{-3}$. In terms of this pair of canonical variables, the Hamiltonian becomes
\be
H = \frac{N}{2} \left[ \frac{\pi^2_{\phi}}{ v} - \kappa^2 vb^2 \right]
\ee
with $\kappa=\sqrt{12\pi G}$ proportional to the Planck length.

Let us now look at the evolution induced by this cosmological Hamiltonian. In particular, we will see that, as introduced in \cite{BenAchour:2019ywl, BenAchour:2018jwq}, the volume forms a closed $\sl(2,\R)$ Lie algebra together with the scalar constraint and the integrated extrinsic curvature. Moreover, we will identify them as the Noether charges for the Mobius transformations.

\subsubsection{The CVH algebra}

Let us look at the evolution of the cosmological variables.
First, since the Hamiltonian does not depend on the scalar field $\phi$ for this simple system, it is then easy to see that  its conjugate momentum $\pi_{\phi}$ is a constant of motion:
\be
\label{com}
\dot{\pi}_{\phi} = \{ \pi_{\phi}, \cH \} =0\,.
\ee
This translates  the continuity equation (\ref{eo2}) to the Hamiltonian framework. 

Now, in order to derive the evolution of the geometrical sector, it is useful to  introduce the trace of the extrinsic curvature $\text{Tr}(\text{K})$, integrated over the fiducial cell living in the spatial hypersurface $\Sigma$:
\be
\ka^2\cC := \int_{V_{\circ}} d^3x \sqrt{\gamma} K = V_{\circ} a^3 \frac{3 \dot{a}}{Na}
=-\f{ \kappa^2}3 \pi_a a
= \kappa^2 bv
\,.
\ee
This phase space function $\cC$ can be identified as the dilatation generator\footnotemark{} on the geometrical sector parametrized by the 3D volume and its conjugated variable, the Hubble factor $b$.
\footnotetext{$\cC$ is actually the dilatation generator on the whole phase space. Notice that it nevertheless does not contain matter contribution. The reason is that the scalar field does not transform as a quasi-primary field under the Mobius transformation, but remains invariant under a conformal transformation \eqref{mobius}, its scaling dimension being $\Delta_{\phi} =0$. }
It turns out that $\cC$ is the speed of the 3D volume in proper time:
\be
\dot v=\frac{\rd v}{\rd \tau} = \{ v, \cH\} = \kappa^2 \cC
\,. 
\ee
Then the time derivative of the dilatation generator is the Hamiltonian itself:
\be
\dot\cC =\{ \cC, \cH\} = -  \cH
\,,\qquad
\dot\cH=0
\,.
\ee
The system of differential equations closes and can be integrated. Moreover, the Poisson brackets of the three variables $v$, $\cC$ and $\cH$ form a closed Lie algebra:
\be
 \{ v, \cH\} = \kappa^2 \cC
\,,\qquad
\{ \cC, \cH\} = -  \cH
\,,\qquad
\{ \cC, v\} =  v
\,.
\ee
This forms an $\sl(2,\mathbb{R})$ Lie algebra, which  we shall refer to as the CVH cosmological algebra. This algebra encodes how the basic physical quantities, the Hamiltonian and the volume, transform under scale transformations. This structure was previously noticed and used in alternative models of quantum cosmology in \cite{BenAchour:2019ywl, BenAchour:2018jwq, BenAchour:2017qpb}.

We can map the $\cC$, $v$ and $\cH$ observables to the usual basis of the $\sl(2,\mathbb{R})$ Lie algebra by
\be
\label{CVHgm}
 k_y = \cC 
 \,, \quad
k_x = \frac{1}{\kappa} \left[ \frac{v}{2\sigma \kappa^2} + \sigma \kappa^2 \cH\right] 
 \,, \quad
 j_z = \frac{1}{\kappa} \left[ \frac{v}{2\sigma \kappa^2} - \sigma \kappa^2 \cH\right] 
 \ee
with the standard commutation relations:
 \be
\{ j_z, k_x\} = k_y
\,, \quad
\{ j_z, k_y\} = - k_x
\,, \quad
\{ k_x, k_y\} = - j_z 
\,,
\nn
\ee
The factor $\sigma$ is an arbitrary parameter, also present in the quantum realization of conformal mechanics \cite{deAlfaro:1976vlx}. It is the proportionality factor in front of the 3d volume and the scalar constraint realized as null generators of the $\sl(2,\mathbb{R})$ algebra:
\be
v=\sigma \ka^3\,\big{(}k_{x}+j_{z}\big{)}
\,,\qquad
\cH=\f1{2\sigma\ka}\,\big{(}k_{x}-j_{z}\big{)}
\,.
\ee
We can compute the Casimir of this $\sl(2,\R)$ algebra:
\be
\label{Csu11}
\mathfrak{C}
=   j^2_z - k^2_x - k^2_y
= - \frac{2}{\kappa^2}v \cH - \cC^2  
=-\frac{\pi^2_{\phi}}{\ka^2}
\,.
\ee
The Casimir is  negative and turns out to be proportional to the constant of motion $\pi^2_{\phi}$ which corresponds to the kinetic energy of the scalar field. Upon quantization of this system, this means that cosmological system gravity plus matter would be described by $\sl(2,\mathbb{R})$ unitary representations representations from  the continuous principal series \cite{BenAchour:2019ywl, BenAchour:2018jwq}. The vacuum case, i.e $\pi_{\phi} =0$, would then correspond to a null representation with vanishing Casimir (up to possible quantum correction due to operator ordering).

We would like to point out that the CVH algebra also applies in the presence of a non-vanishing cosmological constant $\Lambda\ne 0$, in which case the Hamiltonian constraint  becomes a time-like or space-like generator depending on the sign of $\Lambda$ (see Appendix-\ref{app:Lambda}  for more details), and in the presence of anisotropies as in the Bianchi I model.

We will see below that this  $\sl(2,\mathbb{R})$ symmetry is the Hamiltonian counterpart of the invariance of the action under Mobius transformations. More precisely, we will show that CVH generators can be mapped onto the Noether charges for the conformal transformations.

\subsubsection{Charges algebra and their infinitesimal action}
\label{sec:noetherH}

The Noether charges for the Mobius transformations, derived in the previous section \ref{sec:noethervariation}, admit a simple expression in terms of the canonical variables:
\be
\begin{array}{lcl}
Q_{-} & =& \cH \vspace*{1mm}\\
Q_0 & = &\cC + \tau \cH \vspace*{1mm}\\
Q_{+} & = & 2\ka^{-2} v -  2\tau \cC - \tau^2 \cH
\end{array}
\ee
These expressions allow to compute the Poisson brackets of the Noether charges and show that they also form a closed $\sl(2,\R)$ algebra:
\be
\{ Q_0, Q_{\pm}\} = \pm Q_{\pm}\,, \qquad \{ Q_{+}, Q_{-}\} = 2Q_0
\,.
\ee
Furthermore, since the $Q$'s are constant of motions, $\dot{Q_{\mu}}=0$, we can read the trajectories for the 3D volume and the integrated extrinsic curvature:
\be
\label{noetherevolution}
\begin{array}{lcl}
\cH(\tau)& =& Q_{-} \vspace*{1mm}\\
\cC(\tau) & = & Q_{0} - \tau Q_{-} \vspace*{1mm}\\
v(\tau) & = & \f{\ka^2} 2\big{[}Q_{+} +  2\tau Q_{0} - \tau^2 Q_{-}\big{]}
\end{array}
\ee
In particular, the Noether charges are the initial conditions, i.e. the values of $v$, $\cC$ and $\cH$ at $\tau =0$. For physical trajectories of the Einstein-scalar system, the scalar constraint vanishes, $\cH=0$, and we are in the special case where $\cC$ is also a constant of motion.

Finally, we would like to show that the Noether charges indeed generate the Mobius transformations \eqref{mobius}. It is straightforward to compute their Poisson brackets with the basic variables and check that they reproduce the infinitesimal variations of the fields under conformal transformations derived earlier in section  \ref{sec:noethervariation}.
Let us focus on the scale factor $a$; we obtain:
\begin{align}
\epsilon \{ Q_{-} , a \} & = - \epsilon \dot{a}=\delta^t_{\eps}a
\\
\epsilon \{ Q_{0} , a \} & = \frac{\epsilon a}{3} - \epsilon \tau \dot{a} =\delta^d_{\eps}a
\\
\epsilon \{ Q_{+} , a \} & = - \frac{2\epsilon}{3}\tau a+\epsilon \tau^2 \dot{a} =\delta^s_{\eps}a
\end{align}
and  recovers  the infinitesimal variations respectively under  translation (\ref{itrans}), dilatation (\ref{idil}) and special conformal transformation (\ref{ist}) as described in the previous section.

This shows how the CVH Lie algebra is directly related to the invariance of the cosmological action under conformal transformations of the metric realized as Mobius transformations in proper time.

\subsection{Cosmological trajectories from $\SL(2,\R)$ Flow}
\label{evolution}

Since the Poisson brackets of  $v$, $\cC$ and $\cH$ form a  closed  $\sl(2,\R)$ Lie algebra, it is possible to exponentiate this CVH algebra and derive the flow generated by these three observables as flows in the Lie group $\SL(2,\R)$. In particular, as shown in \cite{BenAchour:2019ywl}, this allows to exponentiate the scalar constraint and integrate the evolution in proper time in terms of $\SL(2,\R)$ group elements, leading to a group theory description of the cosmological trajectories.

To this purpose, it is convenient to use the representation of $\SL(2,\R)$ group elements as $\SU(1,1)$ matrices. Following \cite{BenAchour:2019ywl}, we introduce the Hermitian 2$\times$2 matrix formed by the $\sl_{2,\R}$ generators:
\be
\label{M}
M=\mat{cc}{j_{z} & k_{-}\\ k_{+} & j_{z}}
\qquad\textrm{with}\quad
k_{\pm}=k_{x}\pm i k_{y}
\,,
\qquad
\left|\begin{array}{lcl}
\{j_{z},k_{\pm}\}&=&\mp i k_{\pm}
\vspace*{1mm}\\
\{k_{+},k_{-}\}&=&2i j_{z}
\end{array}
\right.
\ee
Writing linear combinations of $v$, $\cC$ and $\cH$ as $\su(1,1)$ Lie algebra vectors,  $\veta\cdot\vcJ=\eta_{z}j_{z}-\eta_{x}k_{x}-\eta_{y}k_{y}$, the  Poisson brackets can be expressed in a compact manner as a matrix multiplication:
\be
\{\vcJ,M\}=\f i 2\,\big{(}
\vlambda M -M \vlambda^{\dagger}
\big{)}\,,
\ee
in terms of Lorentzian Pauli matrices,
\be
\lambda_{z}=\mat{cc}{1 & 0 \\ 0 & -1}
\,,\,\,
\lambda_{x}=\mat{cc}{ 0 & 1 \\ -1 &0}
\,,\,\,
\lambda_{y}=\mat{cc}{ 0 & -i \\-i &0}
\,.
\nn
\ee
These matrices square to the identity, $\lambda_{z}^{2}=\id$ but $\lambda_{x}^{2}=\lambda_{y}^{2}=-\id$, and satisfy the $\su(1,1)$ commutation relations:
\be
[\lambda_{z},\lambda_{x}]=2i\lambda_{y}
\,,\quad
[\lambda_{x},\lambda_{y}]=-2i\lambda_{z}
\,,\quad
[\lambda_{y},\lambda_{z}]=2i\lambda_{x}
\,.
\nn
\ee
This formulation leads to a simple formula for the exponentiated action of Lie algebra vectors on  the matrix $M$:
\be
\label{conjugation}
e^{\{\veta\cdot\vcJ,\,{\bullet}\,\}}\,M=GMG^{\dagger}
\quad\textrm{with}\,\,
G=e^{\f i2 \veta\cdot\vlambda}\in\SU(1,1)
\,.
\ee
We apply this formula to the Hamiltonian flow of the scalar constraint $e^{-\tau\{\cH,\cdot\}}$. Since $\cH=\f1{2\ka\sigma}(k_{x}-j_{z})$ is a null generator in $\sl(2,\R)$, we obtain null $\SU(1,1)$ transformations:
\be
\cG_{\tau}=e^{-i\f \tau{4\ka\sigma}(\lambda_{x}-\lambda_{z})}=\id-i\f \tau {4\ka\sigma}(\lambda_{x}-\lambda_{z})
\,,
\ee
where $(\lambda_{x}-\lambda_{z})$ is a nilpotent matrix with vanishing square. Acting by conjugation with this group element on the generator matrix, $M(\tau)=\cG_{\tau} M^{(0)}\cG_{\tau}^{-1}$  gives the time evolution of the $\su(1,1)$ generators\footnotemark{}, from which we get the trajectories of the cosmological observables in proper time (or equivalently in the gauge $N=1$):
\be
\label{dynamicsCQM}
\left|\begin{array}{ccl}
\cH(\tau)
&=&
\cH^{(0)}\,,
\\
\cC(\tau)
&=&
\cC^{(0)}-\tau\cH^{(0)}\,,
\\
v(\tau)
&=&
v^{(0)}+\tau\ka^2\cC^{(0)}-\f{\tau^{2}\ka^2}2\cH^{(0)}\,,
\end{array}\right.
\ee
\footnotetext{
The evolution of the $\sl(2,\R)$ generators under the Hamiltonian flow generated by the scalar constraint $\cH$ is quadratic in the flow parameter $\eta=\tau/2\ka\sigma$:
\be
\mat{cc}{j_{z}(\eta) & k_{-}(\eta)\\ k_{+}(\eta) & j_{z}(\eta)}
\,=\,
e^{-i\f\eta2(\lambda_{x}-\lambda_{z})}
\mat{cc}{j_{z}^{(0)} & k_{-}^{(0)}\\ k_{+} ^{(0)}& j_{z}^{(0)}}
\Big{(}
e^{-i\f \eta2(\lambda_{x}-\lambda_{z})}
\Big{)}^\dagger
\,,\quad
\left|\begin{array}{ccl}
j_{z}(\eta)
&=&
j_{z}^{(0)}+\eta k_{y}^{(0)}-\f {\eta^{2}}2(k_{x}^{(0)}-j_{z}^{(0)})
\\
k_{x}(\eta)
&=&
k_{x}^{(0)}+\eta k_{y}^{(0)}-\f {\eta^{2}}2(k_{x}^{(0)}-j_{z}^{(0)})
\\
k_{y}(\eta)
&=&
k_{y}^{(0)}-\eta (k_{x}^{(0)}-j_{z}^{(0)})
\end{array}\right.
\nn
\ee
}
where $v^{(0)}$ denotes the initial condition for the volume at $\tau=0$ and so on.
This allows to recover the exact same formula for the cosmological trajectories \eqref{noetherevolution} as we derive using the conformal Noether charges in the previous section \ref{sec:noetherH}. Indeed the Noether charges $Q_{\pm}$ and $Q_{0}$ are identified as the initial conditions $\cH^{(0)}$, $\cC^{(0)}$ and $v^{(0)}$, and are thus clearly constant of motions. Then the expressions for $\cH(\tau)$, $\cC(\tau)$ and $v(\tau)$ are the realization of evolving constants of motion\footnotemark{} in the present cosmological setting.
\footnotetext{
The evolving cosmological observables, generated by the exponentiated Poisson action of the scalar constraint $\exp{[}-\tau\{\cH,\cdot\}{]}$, still form a $\sl(2,\R)$ Lie algebra, just as the original phase space variables $v$, $\cC$ and $\cH$. We refer to it as the dynamical CVH algebra for FLRW cosmology:
\be
\left|
\begin{array}{lcl}
v_{\tau}&=&v+\tau\ka^2\cC-\f{\tau^2\ka^2}2\cH
\\
\cH_{\tau}&=&\cH
\\
\cC_{\tau}&=&\cC-\tau\cH
\end{array}
\right.
\,
\qquad
\left|
\begin{array}{lcl}
\{\cC_{\tau},\cH_{\tau}\}&=&-\cH_{\tau}
\\
\{\cC_{\tau},v_{\tau}\}&=&v_{\tau}
\\
\{v_{\tau},\cH_{\tau}\}&=&\cC_{\tau}
\end{array}
\right.
\,.
\nn
\ee
This explains why both the $v$, $\cC$, $\cH$ observables and the Noether charges $Q_{\pm}$, $Q_{0}$ both form $\sl(2,\R)$  algebras.
}

Restricting our attention to physical trajectories, thus imposing the scalar constraint on initial conditions, $\cH^{(0)}=0$, the Hamiltonian always vanishes and the expression of the trajectories further simplifies.
The dilatation generator becomes a constant of motion along physical cosmological trajectories and gives the speed of the linear growth of the volume:
\be
\cH=\cH^{(0)}=0
\,,\quad
\cC=\cC^{(0)}
\,,\quad
v=\ka^2\cC \tau + v^{(0)}\,.
\ee
The value of the dilatation generator is fixed by the scalar field momentum, up to a sign, by solving the Hamiltonian constraint:
\be
\cC=\pm\f{\pi_{\phi}}{\ka}\,.
\ee
The sign decides if we are in a contracting or expanding phase.
Unfortunately, this method does not provide the evolution of the scalar field. We have to integrate its equation of motion by hand,
\be
\pp_{\tau}\phi=\{\phi,\cH\}=\f{\pi_{\phi}}{v}=\f{\pm\,\ka\,\cC}{\ka^2\cC \tau + v^{(0)}}
\,,
\nn
\ee
\be
\phi=\pm\f{1}{\ka}\,\ln\left(\f{\ka^2\cC \tau+v_{0}}{v_{0}}\right)+\phi_{0}
\,.
\ee
We can write it in a deparametrized fashion by getting rid of the time variable, reflecting the invariance of the theory under time reparametrization:
\be
\label{vphitraj}
v=v^{(0)}\,e^{\pm\ka(\phi-\phi^{(0)})}
\,.
\ee
This describes the cosmological evolution in terms of the scalar field $\phi$ used on an internal clock, with two regimes corresponding to contracting or expanding trajectories.


To summarize, the CVH algebra formed by the 3D volume $v$, the integrated extrinsic curvature $\cC$ and the scalar constraint $\cH$, endows the cosmological phase space and trajectories with a $\SL(2,\R)$ group structure generated by the Noether charges resulting from the invariance of the original action under Mobius transformations in the proper time. This $\sl(2,\R)$ structure is a powerful tool at both classical and quantum level. Indeed, it entirely characterizes the model. In particular, preserving this $\sl(2,\R)$ structure and thus the conformal invariance at the quantum level will provide a stringent criteria to fix all quantization ambiguities, as we will see in section \ref{sec:quantization}.

\subsection{AdS${}_{2}$ as the cosmological phase space}

Before closing this section devoted to the classical theory, we would like to stress one more interesting side product of the $\sl(2,\mathbb{R})$ structure of the cosmological system. We have seen in Section-\ref{DFF} that the cosmological action can be recast into the dAFF model \cite{deAlfaro:1976vlx}. This $\text{CFT}_1$ has known a renewed interest over the last years with respect to the low dimensional $\text{AdS}_2/\text{CFT}_1$ correspondence as a potential candidate for the holographic dual of the two dimension gravity side. From this relation, one could wonder if there is any relationship between our cosmological model and the $\text{AdS}_2$ geometry. As we are going to see, a rather intriguing relationship can indeed be established at the level of the phase space, which relates the cosmological singularity to the null infinity of $\text{AdS}_2$. While we only briefly mention this link, it would be interesting to investigate it further in the future.

\subsubsection{AdS${}_{2}$ geometry}

Let us review the basic properties of the $\text{AdS}_2$ geometry. Using global coordinates  $ t\in \mathbb{R}$ and $r\in \mathbb{R}^{+}$, the $\text{AdS}_2$ metric reads
\be
\label{ads2metric}
\rd s^2 = - \left( 1+ \frac{r^2}{\ell^2}\right) \rd t^2 + \left( 1+ \frac{r^2}{\ell^2}\right) ^{-1}\rd r^2
\,,
\ee
where $\ell$ is the AdS curvature radius.
Contrary to higher dimensional {AdS} spaces, this 2d version has two time-like boundaries. Indeed, taking $r \rightarrow + \infty$, the $g_{rr}$ metric component vanishes, leaving a 1d time-like line. Then, at $r=0$, the geometry reduces to a Minkowski-like line element and at fixed $r$, this provides again a 1d time-like line.

The geodesic equations gives the propagation of test particles on this 2d geometry and are easily written in terms of the proper time (or affine parameter in the massless case) $\tau$:
\be
\frac{\rd t}{\rd\tau} = \cE \left( 1+ \frac{r^2}{\ell^2}\right)^{-1}
\qquad\textrm{and}\qquad
\left|\begin{array}{ll}
\frac{\rd^2r}{\rd\tau^2} + \frac{r}{\ell^2}  =0 \qquad&\text{for massive particles }  \vspace*{2mm}\\
\frac{\rd^2r}{\rd\tau^2}  = 0 \qquad &\text{for massless particles }
\end{array}
\right.
\,,
\ee
where the energy $\cE>0$ is an arbitrary constant.
The solutions, with initial condition $r=0$ at $\tau=0$, are given by
\be
\label{geodesic}
\left|\begin{array}{ll}
r(\tau)  = \ell \sqrt{\cE^{2}-1}\,\sin{\left( \ell^{{-1}}{\tau}\right)} \qquad &\text{for massive particles}\vspace*{2mm}\\
r(\tau)  = \cE\tau \qquad & \text{for massless particles}
\end{array}
\right.
\ee
So, only massless particles following light-like geodesics can propagate towards arbitrary large radius $r$, while massive particles following time-like geodesics always follow bounded trajectories and cannot approach the time-like boundary at asymptotic $r\rightarrow+\infty$.

Furthermore, an efficient method to explore the various foliations and parametrizations of the  $\text{AdS}_2$ geometry is to start with its embedding  in the three-dimensional Minkowski space. Calling the 3D coordinates $\left(X_{0}, X_1, X_2\right)$, the $\text{AdS}_2$ geometry is defined as the one-sheet hyperboloid:
\be
-X^2_{0} + X^2_1 + X^2_2 = \ell^2
\,,
\ee
with the induced metric inherited from the flat 3D metric, $\rd s^{2}=\rd X_{0}^{2}-\rd X_{1}^{2}-\rd X_{2}^{2}$.
The standard form of the metric (\ref{ads2metric}) is obtained by using the coordinates
\begin{align}
X_0 & = \ell \sinh{\rho}=r\,, \\
X_1 & =  \ell \cosh{\rho} \sin\f t{\ell}=\ell\sqrt{1+\f{r^{2}}{\ell^{2}}} \,\sin\f t{\ell}\,,\nn\\
X_2 & =\ell \cosh{\rho} \cos\f t{\ell}=\ell\sqrt{1+\f{r^{2}}{\ell^{2}}} \,\cos\f t{\ell}\,,\nn
\end{align}
Another useful coordinates system $(\eta,u)$ with $u>0$ corresponds to the Poincar\'e patch and is defined by
\begin{align}
X_0 & = \frac{1}{u} \left[ 1+ u^2 \left( \ell^2 - \eta^2\right) \right]\,,\\
X_1 & =\frac{1}{u} \left[ 1- u^2 \left( \ell^2 + \eta^2\right) \right]\nn\,, \\
X_2 & = \ell u \eta\,, \nn
\end{align}
leading to the line element
\be
\label{ads2P}
\rd s^2 = \ell^2 \left( - u^2 \rd \eta^2 + \frac{\rd u^2}{u^2} \right)
\,.
\ee
Using the coordinate $z = 1/u$ instead, it takes the simpler form of a conformally rescaled two-dimensional Minkowski space-time:
\be
\label{ads2PP}
\rd s^2 = \frac{\ell^{2}}{z^2}\,\left(- \rd \eta^2 + \rd z^2\right)
\,.
\ee

\subsubsection{AdS${}_{2}$ as the phase space metric}
\label{ads2}

Having review the basic properties of the $\text{AdS}_2$ geometry, we now show that the cosmological phase space can be endowed with the $\text{AdS}_2$ Lorentzian metric and discuss the interpretation of the physical cosmological trajectories and the  singularity from the $\text{AdS}_2$ point of view. 

The cosmological phase space for FRW cosmology coupled to a homogeneous scalar field is a four-dimensional  manifold parametrized by the four variables $(\phi,\pi_{\phi},b,v)$. Putting the scalar field $\phi$ aside and working at fixed scalar field momentum $\pi_{\phi}$, we have parametrized the two-dimensional gravitational sector $(b,v)$ using the triplet of variables $(\cC,v,\cH)$ satisfying the quadratic condition:
\be
-\f2{\ka^2}v \cH - \cC^2  
=
-\frac{\pi^2_{\phi}}{\ka^2}
\,.
\ee
Moreover, we have shown that this triplet of variable generates $\SL(2,\R)$ group transformations on the cosmological phase space. Mapping this triplet of variables with the $\sl_{2}$ generators in the standard basis,
\be
v=\sigma \ka^3\,\big{(}k_{x}+j_{z}\big{)}
\,,\qquad
\cH=\f1{2\sigma\ka}\,\big{(}k_{x}-j_{z}\big{)}
\,,\qquad
\cC=k_{y}
\,,
\nn
\ee
the quadratic condition above is simply understood as the Casimir equation for the $\sl(2,\R)$ algebra. 

We can write this in terms of metrics on the phase space. We start with the flat 3D metric:
\be
\rd s^2=\rd j_{z}^2-\rd k_{x}^2-\rd k_{y}^2
\,,
\ee
and impose the Casimir equation $j_{z}^2-k_{x}^2-k_{y}^2=-\ell_{\text{AdS}}^2$ with the hyperboloid radius given by $\ell_{\text{AdS}}=\pi_{\phi}/\ka$.
The correspondence with the \ads{} parametrizations is simply given  by $X_{0}=j_{z}$, $X_{1}=k_{x}$ and $X_{2}=k_{y}$, from which we can deduce the relation with $v$ and $b$.
For instance, re-introducing the variables $v=j_{z}+k_{x}$ and $b=v^{-1}k_{y}$, valid on the half-\ads{} patch with $v>0$,  gives the 2D metric:
\be
\rd s^2=\f{\ell_{\text{AdS}}^2}{v^2}\rd v^2-v^2\rd b^2
\,.
\ee
which is nothing else than the $\text{AdS}_2$ metric\footnotemark{} presented in (\ref{ads2P}).
\footnotetext{We can compute its Killing vectors and check that this is a maximally symmetric 2D manifold with three Killing vectors:
\be
\pp_{b}
\,,\quad
v\pp_{v}-b\pp_{b}
\,,\quad
2bv\pp_{v}-\left(b^2+\f {\ell_{\text{AdS}}^2}{v^2}\right)\pp_{b}
\,,
\nn
\ee
which we recognize as the Hamiltonian vector fields for the three observables respectively $v$, $\cC$ and $\cH$.}
The $b$-axis at vanishing volume $v=0$ represents the cosmological singularity, but also defines the configuration space of a quantization in terms of wave-functions $\Psi(b)$.

Different coordinates choices allow to provides alternative intuitive picture of the cosmological singularity\footnote{Another interesting compactification of the \ads{} metric is given by the light-cone variables (see e.g. \cite{Axenides:2013iwa} for a review of various parametrization of the \ads{} space-time). These $u_{\pm}$ coordinates are the ones that transform under Mo\"ebius transformations:
\be
u_{\pm}=\f{k_{x}\pm j_{z}}{\ell-k_{y}}
\,,\quad
u_{+}=\f{v}{\f{\pi_{\phi}}{\ka}-\cC}
\,,\quad
u_{-}=\f{2\cH}{\f{\pi_{\phi}}{\ka}-\cC}
\,,
\nn
\ee
\be
\f{j_{z}}{\ell}=\f{u_{+}-u_{-}}{1+u_{-}u_{+}}
\,,\,\,
\f{k_{x}}{\ell}=\f{u_{+}+u_{-}}{1+u_{-}u_{+}}
\,,\,\,
\f{k_{y}}{\ell}=\f{u_{-}u_{+}-1}{1+u_{-}u_{+}}
\,.
\nn
\ee
In these variables, the metric takes an extremely simple form:
\be
\rd s^2=\f{-4\rd u_{+}\rd u_{-}}{(1+u_{-}u_{+})^2}
\,,
\nn
\ee
which is invariant under the Mo\"ebius transformation:
\be
u_{+}\longmapsto \f{\alpha u_{+}+\beta}{\gamma u_{+}+\delta}
\,,\quad
u_{-}\longmapsto \f{\delta u_{-}-\gamma }{-\beta u_{-}+\alpha}
\,
\nn
\ee
\be
M=\mat{cc}{\alpha & \beta \\ \gamma  & \delta}\in\SL(2,\R)
\,,\quad
{^t}M^{-1}=\mat{cc}{\delta & -\gamma  \\ -\beta & \alpha}
\,.
\nn
\ee
These $\SL(2,\R)$-action is generated by the Poisson brackets with our triplet of observables $(v,\cC,\cH)$:
\be
\left|
\begin{array}{lcl}
\{v,u_{+}\}
&=&
-u_{+}^2
\\
\{2\cH,u_{+}\}
&=&
1
\\
\{C,u_{+}\}
&=&
u_{+} 
\end{array}
\right.
\,,\quad
\left|
\begin{array}{lcl}
\{v,u_{-}\}
&=&
-1
\\
\{2\cH,u_{-}\}
&=&
u_{-}^2
\\
\{C,u_{-}\}
&=&
-u_{-} 
\end{array}
\right.
\,.
\nn
\ee
This pair of light-cone variables would be especially relevant when investigating the possible reformulation of the $\SL(2,\R)$ symmetry of FLRW cosmology in the context of the \ads/CFT${}_{1}$ correspondence.}. Among the different coordinates choice, the Poincar\'e patch is defined by switching $v$ into $v^{-1}$ making explicit that the AdS${}_{2}$ metric is conformally equivalent to the 2d Minkowski metric:
\be
\rd s^2=\f{\ell_{\text{AdS}}^2}{z^2}\,
\big{[}
\rd z^2-\rd B^2
\big{]}
\,,
\qquad\textrm{with}\quad
z=\frac{\ell_{\text{AdS}}^2}{ v}
\,,\quad
B=\ell_{\text{AdS}} b
\,.
\ee
In these Poincar\'e coordinates, the cosmological singularity $v\rightarrow 0$ is mapped to the boundary of $\text{AdS}_2$ at null infinity.
%

So the gravitational phase space at fixed matter conjugate momentum $\pi_{\phi}$ can be identified to the Lorentzian  1+1-dimensional \ads{} manifold with constant curvature, defined as the group quotient $\SL(2,\R)/\SO(1,1)$, with the  \ads{} geodesics corresponding to the flows generated by $\SL(2,\R)$ group elements.
More precisely, the geodesics are drawn by the $\SL(2,\R)$ action generated by exponentiating the Hamiltonian flow of $\sl(2,\R)$ algebra vectors $\vec{\eta}\cdot\vcJ\equiv\,\eta_{\cH}\cH+\eta_{v}v+\eta_{\cC}\cC$.
The first term generates time evolution according to the scalar constraint $\cH$ and $\eta_{\cH}$ is interpreted as the lapse. The second term coupling $\eta_{v}$ is proportional to the cosmological constant, while the third term generates scale transformations.

\subsubsection{From AdS${}_{2}$ geodesics to physical cosmological trajectories}


Endowing the cosmological phase with the $\text{AdS}_2$ metric leads to a mapping of  the  cosmological trajectories generated by the scalar constraint to geodesics of $\text{AdS}_2$.
First focusing on the case of a vanishing cosmological constant $\Lambda=0$, the  scalar constraint is a null generator in the $\sl(2,\R)$ algebra, so  physical  trajectories for such universe should correspond to geodesics of photons on $\text{AdS}_2$ as given by \eqref{geodesic}.
Another way to derive this is to see that  the scalar constraint $\cH=0$ implies that  $v^{2}b^{2}=\ell_{\text{AdS}}^2$, or equivalently $z=B$ using the Poincar\'e coordinates, in which case the line element vanishes  $\rd s^2 \big{|}_{z=B} = 0$.
These geodesics are at $\cC=k_{y}=X_{2}$ constant equal to $\ell_{\text{AdS}}$. This null ray $z=B$ becomes $r=\ell\tan(\ell^{-1}t)$ in the original $\text{AdS}_2$ parametrization \eqref{ads2metric}.
Then these physical cosmological trajectories follow the flow of $\cH$ up to the cosmological singularity at  null infinity in $\text{AdS}_2$.

It is  tempting to view a nearly singular universe as a massless test field approaching the null-infinity of $\text{AdS}_2$.
From this perspective, it would be interesting to see whether numerous investigations and results concerning the physics near the boundary of $\text{AdS}_2$ (among which chaos properties) could be relevant to discuss the regime where a contracting universe approaches the singularity and the fate of the cosmological singularity.

\medskip

Let us now consider the case $\Lambda \neq 0$. As shown in Appendix-\ref{app:Lambda}, depending on whether $\Lambda <0$ or $\Lambda >0$, the scalar constraint can be re-written as a space-like, (respectively time-like) generator in the $\sl(2,\R)$ algebra. Following the previous discussion, the phase space trajectories of such AdS or dS flat cosmologies would correspond to space-like or time-like geodesics on the $\text{AdS}_2$ phase. It is interesting to notice that, because of the infinite potential barrier in $\text{AdS}_2$, such geodesic can never reach the boundary of $\text{AdS}_2$. This fits well with the fact that AdS or dS flat cosmologies are actually free of singularity, being constant curvature spacetimes.

\section{Quantum Cosmology and CFT correlators}
\label{sec:quantization}

We turn now to the quantum theory. Different approaches can be used to perform the quantization of this simple cosmological system. In the following, we shall focus on the canonical Wheeler-de Witt (WdW) quantization scheme and revisit the quantization procedure to implement the newly found conformal symmetry. With this simple quantization, we then show how the standard $\text{CFT}$ correlators emerge in quantum cosmology as the overlap for the evolution of coherent wave-packets. This last proof will be developed in close analogy with \cite{Chamon:2011xk}. 

\subsection{Wheeler-de Witt $\SL(2,\mathbb{R})$ quantization}
\label{sec:wdw}

The construction of the quantum theory is straightforward.
The Hamiltonian scalar constraint is quantized into the Wheeler-de Witt (WdW) differential equation, which is a Klein-Gordon-like equation  relating the variation of the wave function with respect to the gravitational degrees of freedom and the matter degrees of freedom. 
It is a timeless equation (with no explicit time derivative) reflecting the well-known problem of time occurring for systems invariant under time-reparametrizations: the time coordinate disappears in the quantum theory, one then has to reconstruct a time flow, or internal clock, from the timeless observables and operators.

As any canonical quantization scheme, the definition of the quantum theory faces ambiguities. However, the unexpected conformal symmetry of the FLRW action provides a useful criteria to narrow these quantization ambiguities. Below, we revisit the standard WdW quantization in the light on this new invariance of FLRW cosmology under Mobius transformations. This provides a preferred factor ordering and leads to physical wave-functions of the universe solution to the WdW equation and provided with a physical inner product.

Let us start with the FLRW phase space, provided with its canonical brackets,
\be
\{ b, v\} = 1 \qquad \{\phi, \pi_{\phi} \} = 1 
\ee 
and the scalar constraint for flat FLRW cosmology coupled a massless scalar field:
\be
\cH=\f{\pi_{\phi}^2}{2v}-\f{\ka^2}{2}vb^2
\,,
\ee
with the Planck length (up to a numerical factor) $\ka=\sqrt{12\pi G}$. Let us quantize this in the $(v,\phi)$-polarization i.e. using wave-functions $\Psi(v,\phi)$ depending on the volume $v$ and scalar field $\phi$. The canonical variables are raised to differential operators:
\be
\begin{array}{lcl}
\hat{v} \Psi&=&v \Psi 
\,,\vspace*{1mm}\\
\hat{b} \Psi&=&i\pp_{v} \Psi
\,,\vspace*{1mm}\\
{[}\hat{b},\hat{v}{]}&=&i
\,,
\end{array}
\qquad\qquad
\begin{array}{lcl}
\hat{\phi} \Psi&=&\phi \Psi 
\,,\vspace*{1mm}\\
\what{\pi}_{\phi} \Psi&=&-i\pp_{\phi} \Psi
\,,\vspace*{1mm}\\
{[}\hat{\phi},\hat{\pi}_{\phi}{]}&=&i
\,.
\end{array}
\ee
The next step is to quantize the three observables $\cC$, $v$ and $\cH$ and find a suitable factor ordering ensuring that the CVH operators still form a closed $\sl(2,\mathbb{R})$ Lie algebra, ensuring that the algebraic structure of the conformal Noether charges is carried to the quantum level. 
The subtle technical point is the inverse volume factor $\tfrac1v$ in the scalar constraint. In order to avoid that this becomes an issue, we work with wave-functions on the positive real half-line $v\in]0,+\infty[$. The inverse volume $v^{{-1}}$ becomes a legitimate quantum operator acting by multiplication, but we give up exponentiating the operator $\hat{b}$ into finite translations on the real line.

A remark is in order before going any further.
The restriction of the volume to the half-line $\mathbb{R}^{\ast +}$ is natural as we switch our focus from the canonical observable $b$, which generates translations in $v$, to the dilatation generator $vb$, which generates rescalings of $v$. It is interesting to note that switching between momentum variable and dilatation generator is the simplest way to derive the Unruh radiation \cite{Arzano:2018oby} and that this suggests a shortcut to deriving the temperature associated to the FLRW space-time.
We should also mention that restricting to $v\in\mathbb{R}^{\ast +}$ is not a trick to avoid the singularity. Indeed, it does not remove the singularity since divergence and accumulation can still occur as $v\rightarrow 0^{+}$. The absence of singularity would amount to showing that no semi-classical wave packet can evolved such that it ends up peaked around $v \simeq \epsilon$ with $\epsilon \rightarrow 0^{+}$. We underline that the goal here is not to discuss the fate of the cosmological singularity, but to identify a Wheeler-de Witt quantization of the FLRW cosmology compatible with the $\sl(2,\R)$ structure and study its reformulation in CFT terms.
Let us nevertheless remind that the Wheeler-de Witt quantization is typically  not enough to capture the singularity resolution, and more refined quantization schemes have to be used to discuss the quantum UV completion of gravitational system. For instance, \cite{BenAchour:2019ywl} discusses a polymer quantization of FLRW cosmology, preserving the conformal symmetry at quantum level and leading to a resolution of the singularity.
%

\medskip

Let us now quantize the three observables $\cC$, $v$ and $\cH$ and in particular properly define a Hamiltonian constraint operator $\widehat{\cH}$:
\be
\label{repcvh}
\hv
=
v
\,,\qquad
\hcC
=
i\bigg{[}
v\pp_{v}+\f12
\bigg{]}
\,,\qquad
\hcH
=
-\f{1}{2v}\pp_{\phi}^2+\f{\ka^{2}}2\big{[}
v\pp_{v}^2+\pp_{v}
\big{]}
\,,
\ee
This factor ordering is the unique choice ensuring that the operators are Hermitian\footnotemark{} for the scalar product $\la\Psi|\Psi'\ra=\int \rd v\,\rd\phi\,\bar{\Psi}\Psi'$ and form a closed $\sl_{2}$ Lie algebra as wanted:
\be
\label{sl2comm}
[\hat{\cC},\hat{v}]=+i\hat{v}
\,,\qquad
[\hcC,\widehat{\cH}]=-i\widehat{\cH}
\,,\qquad
[\hat{v},\widehat{\cH}]=+i\ka^{2}\hcC
\,.
\ee
\footnotetext{
The choice of scalar product of course changes the operator ordering ensuring that the operators are Hermitian. For instance, the scalar product $\la\Psi|\Psi'\ra_{n}=\int \tfrac1v\rd v\,\rd\phi\,\bar{\Psi}\Psi'$ leads to the normal ordering for the CVH operators:
\be
\hv_{n}
=
v
\,,\qquad
\hcC_{n}
=
iv\pp_{v}
\,,\qquad
\hcH_{n}
=
-\f{1}{2v}\pp_{\phi}^2+\f{\ka^{2}}2
v\pp_{v}^2
\,,
\nn
\ee
whose commutators also form a closed $\sl(2,\R)$ Lie algebra.
}
The Casimir of this $\sl_{2}$ algebra  is given by the energy-momentum of the scalar field:
\be
\label{sl2cas}
\mathfrak{C}_{\sl_{2}}
=
-\f1{\ka^{2}}\Big{[}
\hat{v}\widehat{\cH}+\widehat{\cH}\hat{v}
\Big{]}
-\hcC^2
=
-\f{\hat{\pi}_{\phi}^{2}}{\ka^{2}}-\f14
\,,
\qquad
[\hat{\pi}_{\phi},\hv]=[\hat{\pi}_{\phi},\hcC]=[\hat{\pi}_{\phi},\hcH]=0
\,.
\ee
The $\f14$ term fits well with the Casimir formula  for the continuous series of unitary $\SL(2,\R)$ representations (see Appendix \ref{sec:sl2rirrep} or \cite{AIHPA_1965__3_1_13_0,ruhl1970lorentz,Kitaev:2017hnr} for details on the representation theory of the Lie group $\SL(2,\R)\sim\SU(1,1)$). 

\medskip

The scalar constraint $\widehat{\cH}\,\Psi=0$ now becomes a 2nd order Wheeler-de Witt equation:
\be
v\pp_{v}^2\Psi
+\pp_{v}\Psi
=
\f{1}{\ka^{2}v}\,\pp_{\phi}^2\Psi
\,.
\ee
%
%
The standard way to solve such a Klein-Gordon-like equation is to separate the variables, $\Psi(v,\phi)=\psi(v)\chi(\phi)$, and diagonalizes the two differential operators in $v$ and in $\phi$ on their own. Diagonalizing the scalar field momentum $\what{\pi}_{\phi}$ with plane waves, we are left with a second order differential equation\footnotemark{} on $\psi(v)$ for the geometrical sector:
\be
\left|\begin{array}{ll}
\chi_{p}(\phi)=e^{i\ka p \phi}
\,,\\
\widehat{\pi}_{\phi}\,\chi_{p}=\ka p\,\chi_{p}\,,
\end{array}
\right.
\qquad\quad
v^{2}\pp_{v}^2\psi
+v\pp_{v}\psi
+p^{2}\psi=0
\,.
\ee
\footnotetext{
The case of a vanishing scalar constraint  $\hcH\Psi=0$ is a singular case of the more general eigenvalue problem $\hcH\Psi=\ka^{2}\mu\Psi$ for $\mu\in\C$.  In the general case, the differential equation reads:
\be
v^{2}\pp_{v}^2\psi
+v\pp_{v}\psi
+(p^{2}-2\mu v)\psi=0
\,.
\nn
\ee
This is  solved by modified Bessel functions $\psi(v)\propto I_{\pm 2ip}(2\sqrt{2\mu v})$. The fact that the ``physical'' case $\mu=0$ is a singular limit begs the question of the relevance of possible energy off-set to the scalar constraint operator $\hcH$ due to quantum corrections.
}
This Euler differential equation is easily solved by complex power laws:
\be
\psi^{\pm}_{p}(v)=v^{\pm ip}=e^{\pm ip\ln v}
\,,\qquad
\Psi_{p}^{\pm}(v,\phi)=\psi^{\pm}_{p}(v)\chi_{p}(\phi)=e^{ip(\ka\phi\pm\ln v)}
\,,\qquad
\widehat{\cH}\,\Psi_{p}^{\pm}=0
\,.
\ee
These plane waves in $\phi$ and $\ln v$ provide a basis for the physical cosmological states solving the Hamiltonian scalar constraint. The two possible signs reflect the existence of the two contracting and expanding phases of the classical cosmological evolution.

The original scalar product is sill-defined on those solutions of the WdW equation. Actually the two integrations over $v$ and $\phi$ appears to be redundant. In this situation, it is understood that one should switch from the kinematical scalar product $\la\Psi|\Psi'\ra=\int \rd v\,\rd\phi\,\bar{\Psi}\Psi'$ to a physical inner product suited to the solutions of the WdW equation. One typically fixes the integration over $v$ or over $\phi$:
\begin{align}
\int \rd v\rd \phi\,\ka\delta(v-v_{0})\,
\overline{\Psi}_{p}^{\eps}\Psi_{p'}^{\eps}
&=\delta(p-p')
\,,\\
\int \rd v\rd \phi\,\f1v\delta(\phi-\phi_{0})\,
\overline{\Psi}_{p}^{\eps}\Psi_{p'}^{\eps}
&=\delta(p-p')
\,,
\end{align}
where we assume that the positive sector spanned by wave-functions $\Psi^{+}_{p}$ and the negative by wave-functions $\Psi^{-}_{p}$ are by definition orthogonal. A less ad hoc prescription based on the classical analysis of symmetries, equations of motion and Dirac observables amounts to the integration over the classical trajectories:
\be
\la\Psi|\Psi'\ra_{phys}^{\eps}=
2\ka\int \rd v\rd \phi\,\delta(ve^{-\eps\ka\phi}-v_{0})\,
\overline{\Psi}(v,\phi)\,\Psi'(v,\phi)
\,.
\ee
The scalar product $\la\cdot|\cdot\ra_{phys}^{+}$ is adapted to the positive modes $\Psi^{+}_{p}$  while the scalar product $\la\cdot|\cdot\ra_{phys}^{-}$ is adapted to the negative modes $\Psi^{-}_{p}$:
\be
\la\Psi_{p}^{\eps}|\Psi_{p'}^{\eps}\ra_{phys}^{\eps}=\delta(p'-p)
\,,
\ee
which does not depend on the arbitrary constant $v_{0}$. Having to distinguish between positive and negatives modes is similar to distinguishing between positive and negative energy modes when solving the Klein-Gordon equation.

Let us conclude with a remark on the action of the CVH operators on the space of physical wave-functions.
Operators such as the volume $\hv$ are now either ill-defined or not Hermitian anymore with respect to  the physical scalar product. This is not a deep problem once we remember that the volume $v$ is not a Dirac observable. In particular, it does not legitimately act on the space of solutions to the Wheeler-de-Witt equation: if $\hcH\,\Psi=0$, then $\hcH\hv \,\Psi$ generally does not vanish, as we can check for all basis solutions $\Psi_{p}^{\eps}$.
On the other hand,  the dilatation operator $\hat{\cC}=i(v\pp_{v}+\f12)$ is still well-defined and Hermitian. This works because $\cC$ is classically a Dirac observable on physical trajectories at $\cH=0$.


\bigskip

This realizes the canonical quantization of FLRW cosmology, respecting the $\sl(2,\R)$ symmetry induced by the invariance under Mobius transformations.
A different perspective of the quantization would be to forget the original variables $v$ and $b$, and to shift the focus entirely on the CVH variables. One would start with the $\sl(2,\R)$ commutators \eqref{sl2comm} between $\hv$, $\hcC$ and $\hcH$, and the $\sl(2,\R)$ Casimir equation \eqref{sl2cas}. One would thus quantize the system directly as an irreducible $\SL(2,\R)$ representation determined by the value of the scalar field momentum $\pi_{\phi}$. The  $\SL(2,\R)$ action on the Hilbert space then gives the integrated flow of the three basic operators  $\hv$, $\hcC$ and $\hcH$. This path described in \cite{BenAchour:2019ywl, BenAchour:2018jwq} is ultimately equivalent to the WdW quantization described above, but it nevertheless provides with the whole toolbox of the standard basis of $\sl(2,\R)$ operators, their action and eigenvectors.
It is actually this purely algebraic point of view that we will take below in the next section in order to connect with the conformal field theory (CFT) framework and realize the CFT 2-point function as a cosmological wave-function overlap.

\subsection{CFT Correlators as Wave-packet Overlap in Quantum Cosmology}
\label{sec:CFT}
It is important to keep in mind the conformal invariance of the action of FLRW cosmology is valid off-shell, without necessarily restricting to ``physical'' trajectories satisfying the scalar constraint $\cH=0$. This symmetry applies to the flow of all observables on the full phase space of the theory. At the quantum level, this means that the conformal Noether charges act on the whole kinematical Hilbert space before imposing the scalar constraint operator. Been able to study the whole space of states, and not focusing only on the solution to the present WdW equation $\hcH\Psi=0$, will be especially relevant when the scalar constraint will acquire extra terms either by adding extra matter field or by adding extra interaction terms such gravity modifications or a potential for the scalar field.

Here we would like to show how the Hilbert space for FLRW cosmology carries a representation of CFT${}_{1}$, and in particular how  CFT correlation functions emerge from the quantum cosmology setting.
%

%
Let us start by describing how quantum cosmology could be recast in CFT terms, following what was done for conformal quantum mechanics \cite{Chamon:2011xk}.
In a conformally invariant field theory, a primary operator would be defined as an operator $\cO_{\Delta}(\tau)$ satisfying the following commutations relation with the conformal generators $\hcH$, $\hcC$ and $\hv$,
\begin{align}
i [\hcH, \cO_{\Delta}] & = \dot{\cO}_{\Delta}\;,  \qquad  i [\hcC, \cO_{\Delta}]  = \Delta \cO_{\Delta} \;, \qquad i [\hv, \cO_{\Delta}]  = 0\,.
\end{align}
Now imagine starting with a normalized and conformally-invariant vacuum state $| \Omega \ra$, which  therefore satisfies
\be
\hcH | \Omega \ra =  \hcC | \Omega \ra = \hv| \Omega \ra =0\,
\ee
then the two-points correlation function of the primary operator $\cO(\tau)$ at two different times $\tau_2$ and $\tau_1$ evaluated on this vacuum state are given by
\begin{align}
\label{TWO}
\la \Omega | \cO(\tau_2) \cO(\tau_1) | \Omega \ra & = \frac{1}{|\tau_2 -\tau_1|^{2\Delta}} 
\,,
\end{align}
and so on the higher order correlators.
As underlined in \cite{Chamon:2011xk}, such a vacuum state can not exist due to the non-null $\sl(2,\R)$ Casimir equation \eqref{sl2cas} and there does not exist such a primary operator. A method was nevertheless outlined how to recover the CFT correlators from  quasi-primary operators acting on non-normalizable and non-$\sl(2,\R)$-invariant states. We follow the same approach here and further show that it  corresponds to looking at the time evolution of the dual of coherent cosmological wave-packets.

\medskip

In the following, in this whole section, we will fix the eigenvalue of $\what{\pi}_{\phi}$ and focus on the dynamics of the geometrical sector of the wave-function. This amounts to considering wave-functions $\Psi(v,\phi)=e^{i\pi_{\phi}\phi}\psi(v)$ and consider all the operators as acting as differential operators in the volume $v$ on the wave-functions $\psi(v)$.

\medskip

Let us consider the operators
\be
\label{opa}
\hA:=\f{\hv}{2\sigma\ka^{3}}-i\hcC
\,,\qquad
\hA^{\dagger}=\f{\hv}{2\sigma\ka^{3}}+i\hcC\,,
\ee
where the dilatation generator $\hcC=-i[\hv,\hcH]$ is the speed of the volume in proper time. Using the expression of $\hv$ and $\hcC$ as differential operators given in \eqref{repcvh}, we can diagonalize the operator $A$ for an arbitrary complex eigenvalue $\Delta\in\C$:
\beq
\hA|\zeta^{(\Delta)}\ra
=
\Delta|\zeta^{(\Delta)}\ra
&\,\,\Leftrightarrow\,\,&
v\pp_{v}\zeta^{(\Delta)}+\left(\f12-\Delta+\f v{2\sigma\ka^{3}}\right)\zeta^{(\Delta)}=0
\nn\\
&\,\,\Rightarrow\,\,&
\zeta^{(\Delta)}(v)\propto v^{\Delta-\f12}e^{-\f{v}{2\sigma\ka^{3}}}
\,.
\eeq
The eigenvalue $\Delta$ actually gives the expectation values of both the volume $\hv$ and the integrated extrinsic curvature $\hcC$, giving the states $\zeta^{(\Delta)}$ the interpretation of coherent wave-packets:
\be
\la \hv\ra_{\Delta}
=
\f{\la \zeta^{(\Delta)}|\hv|\zeta^{(\Delta)}\ra}{\la \zeta^{(\Delta)}|\zeta^{(\Delta)}\ra}
=
2\sigma \ka^3 \,\mathfrak{Re}(\Delta)
\,,\qquad
\la \hcC\ra_{\Delta}
=
-\mathfrak{Im}(\Delta)
\,,
\ee
where $\mathfrak{Re}(\Delta)$ and $\mathfrak{Im}(\Delta)$ stands respectively for the real and imaginary parts of $\Delta\in\C$.
%
%
Note that that states are not necessarily normalizable,
\be
\la \zeta^{(\Delta)}|\zeta^{(\Delta)}\ra
\propto
\int_{0}^{+\infty}
v^{2\mathfrak{Re}(\Delta)-1}\,e^{-\f{v}{\sigma\ka^{3}}}
\,,
\nn
\ee
which converges if and only if $2\mathfrak{Re}(\Delta)>0$, i.e. if the expectation value for the volume is strictly positive $\la \hv\ra_{\Delta}>0$.

We are  interested in the time evolution of those coherent wave-packets:
\be
|\zeta^{(\Delta)}_{\tau}\ra
:=
e^{i\tau\hcH}\,|\zeta^{(\Delta)}\ra
\,,
\ee
and more precisely in the overlap between the same state evolved to different times:
\be
G(\tau_{1},\tau_{2})
:=
\la\zeta^{(\Delta)}_{\tau_{1}}|\zeta^{(\Delta)}_{\tau_{2}}\ra
=
\la\zeta^{(\Delta)}|\,e^{i(\tau_{2}-\tau_{1})\hcH}\,|\zeta^{(\Delta)}\ra
\,,
\ee
which depends only the time difference $(\tau_{2}-\tau_{1})$.

\medskip

We will show below that this quantum state overlap is exactly the CFT${}_{1}$ 2-point function and decrease as $|\tau_{2}-\tau_{1}|^{-1}$ for a specific choice of $\Delta$ defining the CFT${}_{1}$ vaccuum state.
Moreover we will formulate this two-point function as the time correlation of vertex-like operator acting on the  state solving the Hamiltonian scalar constraint $\hcH\,|\Omega\ra=0$. This physical state turns out to play the crucial role of the CFT${}_{1}$ vaccuum state.

The proof relies mostly on the $\sl(2,\R)$ Lie algebra structure of the CVH operators and on their resulting time evolution under the flow generated by the scalar constraint operator $\cH$. Then we choose the highest weight  for the $\sl(2,\R)$ algebra, which will lead to the expected result.

\subsubsection{The two-point function from the $\sl(2,\R)$  structure}

As we have seen in the previous section \ref{sec:wdw}, our specific choice of operator ordering ensures that the classical $\sl(2,\R)$ structure of the CVH observables carries on to the quantum level. The mapping of the CVH algebra,
\be
[\hcC,\hv]=i\hv
\,,\qquad
[\hcC,\hcH]=-i\hcH
\,,\qquad
[\hv,\hcH]=i\ka^{2}\hcC
\,,
\nn
\ee
to the $\sl(2,\R)$ Lie algebra standard basis is a simple linear map:
\be
\hv=\sigma\ka^{3}(\hk_{x}+\hj_{z})
\,,\quad
\hcH=\f1{2\sigma\ka}(\hk_{x}-\hj_{z})
\,,\quad
\hcC=\hk_{y}
\,,\quad
\hk_{x}=\f{\hv}{2\sigma\ka^{3}}+\sigma\ka\hcH
\,,\quad
\hj_{z}=\f{\hv}{2\sigma\ka^{3}}-\sigma\ka\hcH
\,,
\nn
\ee
with the $\sl(2,\R)$ commutators:
\be
[\hk_{x},\hk_{y}]=-i\hj_{z}
\,,\qquad
[\hk_{x},\hj_{z}]=+i\hk_{x}
\,,\qquad
[\hj_{z},\hk_{x}]=+i\hk_{y}
\,.
\nn
\ee
It is convenient to write in terms of the lowering and raising operators of the $\sl(2,\R)$ Lie algebra, in order to connect with the operators $\hA$ and $\hA^{\dagger}$:
\be
\hk_{\pm}=\hk_{x}\pm i\hk_{y}
\,,\qquad
\left|\begin{array}{lcl}
\hk_{-}&=& \hA+\sigma \ka\hcH
\,,\vspace*{1mm}\\
\hk_{+}&=& \hA^{\dagger}+\sigma \ka\hcH
\,,
\end{array}\right.
\qquad
\left|\begin{array}{lcl}
{[}\hk_{+},\hk_{-}{]}&=& -2\hj_{z}
\,,\vspace*{1mm}\\
{[}\hj_{z},\hk_{\pm}{]}&=& \pm\hk_{\pm}
\,.
\end{array}\right.
\ee
This Lie algebra structure is immediate to integrate to  $\SL(2,\R)$ Lie group flows. This gives the same proper time evolution for the quantum operators as the classical trajectories:
\be
\left|\begin{array}{lclcl}
\hcH(\tau)&=& e^{i\tau\hcH}\,\hcH \,e^{-i\tau\hcH}&=&\hcH
\,,\vspace*{1mm}\\
\hcC(\tau)&=& e^{i\tau\hcH}\,\hcC\, e^{-i\tau\hcH}&=&\hcC-\tau\hcH
\,,\vspace*{1mm}\\
\hv(\tau)&=& e^{i\tau\hcH}\,\hv\, e^{-i\tau\hcH}&=&\hv+\tau\ka^{2}\hcC-\f{\tau^{2}}2\ka^{2}\hcH
\,.
\end{array}\right.
\ee
Plugging these expressions in the definition of the operator $\hA$ gives its time evolution. In particular, this shows that $\hA$ and $\hA^{\dagger}$ are thermal evolutions of the $\hj_{z}$ operator, i.e. evolution in imaginary proper time:
\be
\left|\begin{array}{lclcl}
\hA(\tau=+2\sigma\ka i)
&=&
e^{-2\sigma\ka\hcH}\,\hA \,e^{+2\sigma\ka\hcH}&=&\hj_{z}
\,,\vspace*{1mm}\\
\hA^{\dagger}(\tau=-2\sigma\ka i)
&=&
e^{+2\sigma\ka\hcH}\,\hA^{\dagger} \,e^{-2\sigma\ka\hcH}&=&\hj_{z}
\,.
\end{array}\right.
\ee
This allows to map eigenstates of the $\hA$ operators to eigenvectors of the $\sl(2,\R)$ Cartan operator $\hj_{z}$:
\be
\label{zetawavefunction}
\hA|\zeta^{(\Delta)}\ra
=
\Delta|\zeta^{(\Delta)}\ra
\,\,\Leftrightarrow\,\,
\hj_{z}\,|\Delta\ra
=
\Delta\,|\Delta\ra
\qquad\textrm{with}\quad
|\zeta^{(\Delta)}\ra
=
e^{+2\sigma\ka\hcH}\,|\Delta\ra
\,.
\ee

We are now ready to compute the overlap resulting from the time evolution of the eigenstates of $\hA$:
\be
G(\tau_{1},\tau_{2})
:=
\la\zeta^{(\Delta)}_{\tau_{1}}|\zeta^{(\Delta)}_{\tau_{2}}\ra
=
\la\Delta|\,e^{+2\sigma\ka\hcH}e^{i(\tau_{2}-\tau_{1})\hcH}e^{+2\sigma\ka\hcH}\,|\Delta\ra
\,.
\ee
Writing this in terms of a single time label $G(\tau):=\la\zeta^{(\Delta)}|\zeta^{(\Delta)}_{\tau}\ra$, we compute its time derivative:
\beq
\tau\pp_{\tau}G
&=&
i\la\zeta^{(\Delta)}|\tau\hcH e^{i\tau\hcH}|\zeta^{(\Delta)}\ra
=
i\la\zeta^{(\Delta)}|\,\big{(}\hcC-\hcC(\tau)\big{)}\,e^{i\tau\hcH}|\zeta^{(\Delta)}\ra
\nn\\
&=&
i\la\zeta^{(\Delta)}|\,\hcC e^{i\tau\hcH}|\zeta^{(\Delta)}\ra
-
i\la\zeta^{(\Delta)}|\,e^{i\tau\hcH}\hcC |\zeta^{(\Delta)}\ra
\nn\\
&=&
\la\Delta|\,e^{+2\sigma\ka\hcH}\,i\hcC \,e^{i\tau\hcH}e^{+2\sigma\ka\hcH}|\Delta\ra
-
\la\Delta|\,e^{+2\sigma\ka\hcH}e^{i\tau\hcH}\,i\hcC \,e^{+2\sigma\ka\hcH}|\Delta\ra
\,.
\label{Gcorrdef}
\eeq
A few algebraic manipulations\footnotemark{} leads to:
\be
\label{Gcorrform}
\tau\pp_{\tau}G
=
+(\Delta+\bar\Delta)\,G
-\la\Delta|\,\hk_{-}\,e^{(i\tau+2\sigma\ka)\hcH}\,|\Delta\ra
-\la\Delta|\,e^{(i\tau+2\sigma\ka)\hcH}\,\hk_{+}\,|\Delta\ra
\,.
\ee
\footnotetext{
Expressing the dilatation generator  back in terms of the $\hA$ operator gives a simple expression:
\be
i\hcC= \hj_{z}+\sigma\ka\hcH-\hA=\hA^{\dagger}-\hj_{z}-\sigma\ka\hcH\,,
\nn
\ee
which leads to:
\beq
i\hcC \,e^{+2\sigma\ka\hcH}
&=&
-\hA\,e^{+2\sigma\ka\hcH}+e^{+2\sigma\ka\hcH}\hk_{+}
\nn
\,,\\\nn
e^{+2\sigma\ka\hcH}\,i\hcC
&=&
e^{+2\sigma\ka\hcH}\hA^{\dagger}-\hk_{-}e^{+2\sigma\ka\hcH}
\,.\nn
\eeq
%
%
Plugging these into the derivative of the two-point function \eqref{Gcorrdef} leads to the result \eqref{Gcorrform}.
}
Thus assuming that the state $|\Delta\ra$ is a highest weight vector, $\hk_{+}\,|\Delta\ra=0$ leads to the elegant  equation:
\be
\tau\pp_{\tau}G
=
+2\mathfrak{Re}(\Delta)\,G\,,
\ee
which integrates the expected simple behavior for the two-point function:
\be
\label{GsolDelta}
G(\tau)
=
\la\zeta^{(\Delta)}_{0}|\zeta^{(\Delta)}_{\tau}\ra
\propto
{\tau^{2\mathfrak{Re}(\Delta)}}
\,.
\ee
Finally, since the $\sl(2,\R)$ Casimir of the CVH algebra is determined by the scalar field momentum $\pi_{\phi}$, it also determines the lowest weight of the corresponding representation as we review below.

\subsubsection{$\SL(2,\R)$ irreducible representations and lowest weight vector}

The Casimir equation \eqref{sl2cas} leads to a negative $\sl(2,\R)$ Casimir, as expected from the classical theory:
\be
\mathfrak{C}_{\sl_{2}}
=
\hj_{z}^{2}-\hk_{x}^{2}-\hk_{y}^{2}
=
-\f1{\ka^{2}}\Big{[}
\hat{v}\widehat{\cH}+\widehat{\cH}\hat{v}
\Big{]}
-\hcC^2
=
-s^{2}-\f14
\qquad\textrm{with}\quad
s:=\f{\pi_{\phi}}\ka
\,.
\ee
As reviewed in appendix \ref{sec:sl2rirrep}, this determines an irreducible unitary representation of $\SL(2,\R)$  from the principal continuous series. Such representation are labeled by a complex spin $j=-\f12+is$ and the action of the $\sl(2,\R)$ generators on eigenstates of the Cartan element $\hj_{z}$ are:
\begin{align}
\mathfrak{C} | j, \Delta \ra & = j(j+1) | j, \Delta \ra = -\left(s^{2}+\f14\right)\,| j, \Delta \ra
\,,\nn\\
\hj_{z} | j, \Delta \ra & = \Delta | j, \Delta \ra
\,,\nn\\
\hk_{+} | j, \Delta \ra & = (\Delta-j)^{\f12}(\Delta+j+1)^{\f12}\,| j, \Delta +1 \ra
\,,\label{kplus}\\
\hk_{-} | j, \Delta \ra & =  (\Delta+j)^{\f12}(\Delta-j-1)^{\f12}\,| j, \Delta-1 \ra
\,.\nn
\end{align}
As underlined in the thorough analysis  of the eigenstates of all $\sl(2,\R)$ operators in $\SL(2,\R)$  unitary representations presented in \cite{lindblad}, the operator $\hj_{z}$ admits eigenfunctions for all complex eigenvalues $\Delta\in\C$. These are generally non-normalizable states, they live in the space of slowly increasing wave-functions, which is much larger than the Hilbert space of $L^{2}$ wave-functions. The fact that $\hj_{z}$ is self-adjoint means that the eigenfunctions for real eigenvalues $\Delta\in\R$ span the Hilbert space of quantum states and form a decomposition of the identity.

In this context, our proposal is to use the (non-normalizable) highest weight vector $| j, \Delta_{0} \ra$  for the mapping of FLRW cosmology to a CFT${}_{1}$. This vector satisfies $\hk_{+}| j, \Delta_{0} \ra=0$, which determines the eigenvalue:
\be
\Delta_{0}=+j \textrm{ or }-(j+1)
\quad\textrm{i.e.}\quad
\Delta_{0}=-\f12\pm is
\,.
\ee
Let us point out that this leads to an a priori negative expectation value for the volume operator $\la \hv\ra_{\Delta}=-\sigma\ka^{3}<0$ and is actually a non-normalizable state
$\la \zeta^{(\Delta)}|\zeta^{(\Delta)}\ra
\propto
\int_{0}^{+\infty}
v^{-2}\,\exp[\sfrac{-v}{\sigma\ka^{3}}]=\infty$.

Merging with the result of the previous section, this determines the eigenvalue $\Delta$ entering the equation \eqref{GsolDelta} and leads to the expected decreasing behavior for the overlap:
\be
\label{GsolDelta0}
G(\tau)\propto\f1{\tau}\,.
\ee
The divergence at the initial time, as $\tau\rightarrow 0^{+}$, reflects that the state is non-normalizable.
This fits with the results of \cite{Chamon:2011xk} obtained for irreducible unitary $\SL(2,\R)$ representations from the principal discrete series, with positive Casimir. In that case, the scaling dimension depends explicitly on the considered representation. Here in our case, considering irreducible unitary representations from the principal continuous series, with negative Casimir, the representation label $s=\pi_{\phi}/\ka$ does not affect the scaling of the time overlap function.

\subsubsection{Vaccuum state and quasi-primary operator}

We have to identified the CFT${}_{1}$ two-point function $G(\tau_{1},\tau_{2})$ to the overlap of the evolution in proper time $\la \zeta_{\tau_{1}}|\zeta_{\tau_{2}}\ra$. In order to truly understand the mapping of  quantum FLRW cosmology to the CFT framework, we still need to write this correlation in terms of the two-point correlation of an observable insertion evaluated on a vaccuum state.

As for now, we have written the  two-point function as:
\beq
G(\tau_{1},\tau_{2})
=
\la \zeta_{\tau_{1}}|\zeta_{\tau_{2}}\ra
&=&
\la \zeta^{(\Delta_{0})}| \,e^{i(\tau_{2}-\tau_{1})\hcH}\,|\zeta^{(\Delta_{0})}\ra \nn\\
&=&
\la \Delta_{0}|\,e^{+2\sigma\ka\hcH}e^{i(\tau_{2}-\tau_{1})\hcH}e^{+2\sigma\ka\hcH}\,|\Delta_{0}\ra
\,,\nn
\eeq
evaluated on the states:
\be
\hA\,|\zeta^{(\Delta_{0})}\ra=\Delta_{0}\,|\zeta^{(\Delta_{0})}\ra
\,,\qquad
\hj_{z}\,|\Delta_{0}\ra=\Delta_{0}\,|\Delta_{0}\ra
\,,\qquad
\textrm{with}\quad
\Delta_{0}=-\f12\pm i\f{\pi_{\phi}}\ka
\,.
\ee
As a wave-function of the volume $v$, we have already computed the expression of the state $|\zeta^{(\Delta_{0})}\ra$ earlier in \eqref{zetawavefunction}. Choosing $\Delta_{0}=-\f12+ i\ka^{-1}{\pi_{\phi}}$, we get:
\be
\label{eigA}
\zeta^{(\Delta_{0})}(v)
=
v^{-1+i\f{\pi_{\phi}}\ka}e^{-\f v{2\sigma\ka^{3}}}
\,.
\ee
It seems natural to use the physical state solving the scalar constraint as the vaccuum of quantum cosmology. As analyzed in the previous section \ref{sec:wdw} on the Wheeler-de Witt quantization, the physical state reads:
\be
\label{cosmovac}
\hcH\,|\Omega\ra=0
\,,\qquad
\psi_{\Omega}(v)=v^{+i\f{\pi_{\phi}}\ka}\,.
\ee
Clearly the eigenstates of $\hA$ can be obtained from the physical state by the action of the equivalent of a vertex operator in our simple setting:
\be
|\zeta^{(\Delta_{0})}\ra
\,=\,
\hcO_{\sigma}\,|\Omega\ra
\qquad\textrm{with}\quad
\hcO_{\sigma}
=
\hv^{-1} e^{-\f \hv{2\sigma\ka^{3}}}
\,.
\ee
This means that the two-point function is obtained by the insertion of the time-evolved operator $\hcO_{\sigma}(\tau)=e^{i\tau\hcH}\hcO_{\sigma}e^{-i\tau\hcH}$:
\be
G(\tau_{1},\tau_{2})
=
\la \zeta_{\tau_{1}}|\zeta_{\tau_{2}}\ra
=
\la \Omega |\, \hcO_{\sigma}(\tau_{1})\hcO_{\sigma}(\tau_{2})\,| \Omega\ra
\,.
\ee
The time ordered version of this operator correlation fits exactly with the desired CFT two-point function:
\be
\label{finn}
\la \Omega |\, \cT\Big{[}\hcO_{\sigma}(\tau_{1})\hcO_{\sigma}(\tau_{2})\Big{]}\,| \Omega\ra
\propto
\f1{|\tau_{1}-\tau_{2}|}
\,.
\nn
\ee
This establishes the first step of a reformulation\footnotemark{} of FLRW quantum cosmology as a CFT${}_{1}$.
\footnotetext{
We have focused on the $\SL(2,\R)$ symmetry. An interesting remark is that it is possible to identify a whole ladder of charges forming a Virasoro algebra for conformal quantum mechanics, as shown in \cite{Kumar:1999fx, Cacciatori:1999rp, Mignemi:2000cv}. These charges are nevertheless entirely determined by the three $\sl(2,\R)$ Noether charges due to the finite number of degrees of freedom of the theory. If this Virasoro structure can also be realized in FLRW cosmology, this would pave the way to implement unitarily the time-diffeomorhphism at the level of the Hilbert space, providing a first realization of quantum covariance for such very simple gravitational system.
}
As explained in \cite{Chamon:2011xk}, the caveat of this approach is that $\hcO_{\sigma}$ is not a true primary operator. Its commutator with $\hcH$ and $\hcC$ involves non-trivial operators.
However, it realizes the intuitive idea that the vacuum of the theory is the physical state annihilating the Hamiltonian scalar constraint.

More work is required to reformulate the 3-point function and higher correlations in cosmological terms, especially to analyze and get a deeper understanding of the notion of quasi-primary operators. In the meanwhile, the toolbox introduced here, using the $\sl(2,\R)$ algebraic structure, allows quite generally to compute the time evolution of operators and time overlap of states (see e.g. the overlap of thermal dilatation eigenstates presented in appendix \ref{sec:boostedC}).

Finally, we would like to comment that we have not constructed the quantum states over a $\SL(2,\mathbb{R})$-invariant vacuum state but over the cosmological vacuum annihilating the Hamiltonian operator. This can be considered as natural from the perspective of $\SL(2,\mathbb{R})$ as a symmetry but not as a gauge symmetry. For each value of the matter energy-momentum $\pi_{\phi}$, we do work with a unitary representation of the 1D conformal group $\SL(2,\mathbb{R})$, which allows to use the $\sl(2,\mathbb{R})$ algebraic structure to write down differential equations entirely determining  the correlation functions as in a conformal bootstrap. When  $\pi_{\phi}$ vanishes, the pure gravity case does not live in the trivial representation of $\sl(2,\mathbb{R})$ but in a null representation of $\sl(2,\mathbb{R})$ with vanishing quadratic Casimir. Following earlier work on conformal quantum mechanics \cite{deAlfaro:1976vlx}, it is nevertheless possible to construct a $\SL(2,\mathbb{R})$-invariant density matrix, interpreted as a vacuum mixed state with thermal properties.


\section{Discussion}

We have shown that FLRW cosmology, defined as the homogeneous and isotropic symmetry reduced action of the Einstein-Scalar system, enjoys a hidden conformal symmetry. Under Mobius transformation of the proper time, with the scalar factor transforming  as a primary field, the dynamics of General Relativity minimally coupled to a massless scalar field, in such symmetry reduced situation, remains invariant. As expected from the Noether's theorem, this conformal symmetry of the action is related to three conserved charges which are not surface charges but are associated to the bulk region. At the Hamiltonian level, these charges form an $\sl(2,\mathbb{R})$ Lie algebra and generate the infinitesimal translation, dilatation and special conformal transformations. They turn out to provide to initial conditions for the extrinsic curvature $\cC$, the 3D volume $v$ and the Hamiltonian scalar constraint $\cH$. These three phase space functions form the CVH algebra unveiled in \cite{BenAchour:2019ywl, BenAchour:2018jwq, BenAchour:2017qpb} which form the hamiltonian realization of the conformal symmetry. It is worth pointing that it is precisely this CVH structure  which initially, suggested the existence of the hidden conformal symmetry reported here. 

Finally, we have provided a precise mapping with the conformal mechanics introduced by  de Alfaro, Fubini and Furlan (dAFF) \cite{deAlfaro:1976vlx}, showing that the FLRW cosmology action written in proper time coincides with the conformal mechanics action. At the classical level, this shows that the homogeneous and isotropic Einstein-Scalar system is a conformal invariant field theory in one dimension, i.e a $\text{CFT}_1$. 

At the quantum level, the conformal symmetry allows to bootstrap the quantum theory. Let us summarize the main results and underline the two sides of our computation.
\begin{itemize}
\item On one hand, we have shown how to adapt the standard Wheeler-de Witt canonical quantization in order to preserve the $\sl(2,\mathbb{R})$ symmetry. This conformal structure, through the CVH algebra, is a powerful criterion to narrow the quantization ambiguities, such as the factor ordering ones. Moreover, the presence of this conformal structure requires to use of an $\sl(2,\mathbb{R})$ invariant scalar product. 
\item On the other hand, we have shown how the standard form of the two-point CFT correlator can be reproduced in FLRW quantum cosmology. Let us emphasize that that the derivation of the main differential equation (\ref{Gcorrform}) for the two points function relies solely on the $\sl(2,\mathbb{R})$ algebraic structure of the system, and as such, is completely independent from the WdW quantization. The final result (\ref{GsolDelta0}) descends only from this key equation (\ref{Gcorrform}) for the two points function, provided one works with the lowest eigenstate $\hat{k}_{+} | \Delta_0 \ra =0$. It means that one can identify an operator, given by (\ref{opa}), and therefore a state, which reproduces the standard expression of the CFT two-points functions. Then, we have used the WdW quantization to obtain a physical interpretation of this operator and the associated eigenstate, and make a link with the standard mini-superspace approach. Using the WdW $\sl(2,\mathbb{R})$ invariant quantization, we have shown that the eigenstate (\ref{eigA}) can be viewed as a coherent wave packet. Moreover, we have shown how the specific state $|\zeta_{\Delta_0}\ra$ corresponding to $\Delta=\Delta_0 = - \frac{1}{2} \pm is$ is obtained from the cosmological vacuum state $| \Omega \ra$ annihilated by the scalar constraint, given by (\ref{cosmovac}). Therefore, even if once cannot realize the two-points function on a conformally invariant vacuum state as in 2d CFT, we have shown that this correlator is naturally realized as the overlap of cosmological coherent wave packet at time $\tau_2$ and $\tau_1$, which furthermore, can be expressed as the two-points function of a vertex operator $\hat{\cO}(\tau)$ evaluated on the cosmological vacuum state $| \Omega \ra$, given by (\ref{finn}). %
\end{itemize}

This first computation of the two-points function suggests that the results and technics developed for the conformal mechanics at the quantum level, such as the computation of the three-points and four points functions using the conformal structure \cite{Jackiw:2012ur}, might be applicable to FLRW quantum cosmology. It would be interesting to push further the computation of the higher order correlation functions following this bootstrap strategy.

Let us now underline some crucial difference with previous works. As already mentioned in the introduction, it is now well known that test fields propagating on cosmological background, such as de Sitter or FRW backgrounds enjoy some exact or approximative conformal symmetry \cite{ Anninos:2011af, Kehagias:2013xga}. In this work, we have obtained a rather new type of conformal symmetry, as it applies to the dynamical background itself. It would be interesting to understand how the conformal symmetry of the background is translated at the level of the test fields, through the existence of conformal Killing vectors. Indeed, under our Mobius transformation, the cosmolgical metric transforms as (\ref{transfo}), which can be rewritten as a global time-dependent conformal rescaling by an appropriate choice of lapse function. Such transformation might be related to conformal Killing vectors, which also appear to play a crucial role in the conformal properties of test fields. Hence, it would be interesting to investigate deeper the relationship between conformal structure of the background and the test fields.  

Moreover, let us point that the conformal symmetry reported here is realized in the bulk and is not tied to some boundary conditions. The conserved Noether's charges are defined as integration on the bulk, and are therefore quite different from the standard gravitational surface charges discussed in the literature. Whether the present reformulation of FLRW cosmology as a $\text{CFT}_1$ can find a holographic interpretation remains to be shown. If this is possible, this might provides an interesting new path to construct a FLRW/CFT correspondence. See \cite{Hertog:2004rz, Hertog:2005hu, Craps:2007ch} for some proposals concerning holographic quantum cosmology models.

From a more general point of view, the results presented above might also be relevant when trying to construct a phase transition interpretation of quantum cosmology. Indeed, quantum cosmological spacetimes can only emerge through a coarse-graining of the underlying quantum geometry, since most of the degrees of freedom have been integrated out. The idea that quantum cosmological geometries correspond to phase transitions of the underlying quantum geometry, and thus to some fixed point of the renormalization flow, is not new. However, one needs to identify the conformal symmetry associated to this fixed point. Hence, the conformal structure presented here might provide the missing ingredient to construct such scenario. It would be interesting to discuss if one can make sense of this phase transition picture of cosmological spacetime in known models of quantum gravity. An interesting well developed framework to test this idea would be the GFT condensates \cite{Gielen:2013naa, Oriti:2016ueo, Oriti:2016qtz, Oriti:2016acw}. A new input suggested by our work would be to use the so called quintic non-linear Schrodinger equations (NLS) as mean field approximation schemes instead of the Gross-Pitaevskii's one, in order to take into account the newly found conformal symmetry. See \cite{Lidsey:2018byv} for details\footnote{ In this work, another type of conformal structure was found by mapping the cosmological dynamics onto a quintic non-linear Schrodinger equation which enjoys an $\sl(2,\mathbb{R})$ symmetry. However, at this stage, it is hard to make contact with this finding. Indeed, while the conformal symmetry affects the equation of motion in \cite{Lidsey:2018byv}, the conformal symmetry reported here acts at the level of the action. Moreover, it seems that the conformal structure found in \cite{Lidsey:2018byv} is different in that it includes both a self-interacting potential $V(\phi)$ for the scalar field, as well as homogeneously curved backgrounds. In our case, adding these two ingredients would break the CVH algebra and thus the conformal symmetry. Yet it might be that the two structures found here and in \cite{Lidsey:2018byv} have a common ground, but this requires more works to be fully understood. }.

Moreover, the possibility to rewrite the cosmological action as a Schwarzian calls for additional investigations. It would be interesting to make contact with the recent works  \cite{Mertens:2018fds, Blommaert:2018oro, Mertens:2017mtv, Lam:2018pvp, Belokurov:2018aol} concerning the Schwarzian theory. Additionally, the appearance of the $\text{AdS}_2$ metric on the cosmoloigcal phase space begs also for further investigations. In particular, a better understanding of the relationship between $\text{AdS}_2$ geodesics and physical trajectories of the universe in phase space could open interesting directions. For example, provided this structure extends to more complex backgrounds, such as the Bianchi IX model, it would be interesting to discuss the cosmological interpretation of the chaotic behavior investigated in the context of $\text{AdS}_2$ \cite{Haehl:2017pak, Jensen:2016pah, Turiaci:2016cvo}. Interestingly, it might shed some light on the chaotic behavior of the universe near the singularity discussed in various cosmological models, and in particular in the Bianchi IX (or mixmaster) universe. See \cite{Cornish:1997ah, Damour:2002et} for details.

As a final comment, we point out that the conformal structure found here is not tied to the highly symmetric framework of homogeneity and isotropy. The CVH structure have been shown to hold for the Bianchi I model in \cite{BenAchour:2019ywl}, while the deformation of the CVH algebra due to the inclusion of a cosmological constant is discussed in Appendix-\ref{app:Lambda}. A complete investigation of the fate of this conformal structure in a wider set of gravitational systems is mandatory. On-going efforts concerning the spherically symmetric Einstein-Scalar system will be presented in a future work. Moreover, the fate of this symmetry in presence of perturbative inhomogeneities would provide another interesting development. In particular, it is crucial to understand how cosmological perturbations break the newly found conformal symmetry. 

\acknowledgments

The work of J.BA was supported by Japan Society for the Promotion of Science Grants-in-Aid for Scientific Research No. 17H02890.
\newpage
\appendix


\section{Conformal structure of the homogeneous (A)dS universe}
\label{app:Lambda}

For the sake of completeness, we describe how the $\su(1,1)$ structure extends to a non-vanishing cosmological constant. Although this will not be relevant to the rest of the discussion, in particular the cosmological constant does not enter the mapping of quantum cosmology onto conformal quantum mechanics, it seems instructive to take into account the role of the  cosmological constant whenever possible.

Consider again the FLRW metric
\be
\rd s^2 = - N^2(t) \rd t^2 + a^2(t) \delta_{ij} \rd x^i \rd x^j 
\ee
and the symmetry reduced action of the Einstein-Scalar system  with a cosmological constant integrated over a cell of volume $V_{\circ}$
\be
\label{lambdaaction}
\cS[N, a, \dot{a}, \dot{\phi}]
 =
V_{\circ} \int_{\mathbb{R}}N \rd t
 \,\left[
 a^3\frac{\dt{\phi}^2}{2N^2} 
- \frac{3}{8\pi G}\frac{a \dt{a}^2}{N^2} - \frac{ \Lambda}{8\pi G} a^3
 \right]
 \,,
\ee
Let us now show how the CVH algebra is deformed by the presence of the cosmological constant. 
\subsection{The deformed CVH algebra}

Performing the canonical analysis, the flat homogeneous and isotropic cosmology phase space with a non-vanishing cosmological constant is given by a modification of the Hamiltonian scalar constraint:
\be
\label{Hlambda}
\cH_{\Lambda} = \frac{\pi^2_{\phi}}{2v} - \frac{\ka^2}{2} v b^2 + \frac{\Lambda}{3\ka^2} v\,.
\ee
This changes the Poisson-algebra with the volume $v$ and dilatation generator $\cC$. Indeed the cosmological constant comes in front a volume term in the Hamiltonian constraint, which does not scale as the other terms:
\be
\{ v, \cH_{\Lambda}\} = \ka^{2}\cC
\,,\quad
\{ \cC, v\} = v
\,,\qquad
\text{but} \quad \{ \cC, \cH_{\Lambda}\} = - \cH_{\Lambda} + \frac{2\Lambda}{3\ka^2} v
\,.
\nn
\ee
The CVH algebra is still closed but its identification to the $\su(1,1)$ Lie algebra is slightly modified. We keep the same expressions for the dilatation generator and the volume
\be
 \cC = K_y 
 \,, \quad
 v =\sigma\ka^{3}\big{(} k_x + j_z \big{)}
 \,,
 \nn
\ee
but we change the relation between the Hamiltonian constraint and the $\su(1,1)$ generators:
\begin{equation}
\label{CosmoConst}
\cH_{\Lambda} = \f1{2\sigma\ka}
\Bigg{[}
\left(1 + \frac{2\Lambda\sigma^{2}\ka^2}{3}\right) k_x - \left(1 - \frac{2\Lambda\sigma^{2}\ka^2}{3}\right) j_z
\Bigg{]}
\,.
\end{equation}
So the Hamiltonian constraint is not anymore a null-vector of the $\sl(2,\R)$ Lie algebra. It becomes a space-like vector generating boosts when $\Lambda >0$, or a time-like vector generating rotations when $\Lambda <0$.

 The dilatation $\cC = vb$ is no more a constant of motion along the Hamiltonian trajectories. The replacing constant of motion is:
\be
\left\{v\sqrt{b^{2}-\f{2\Lambda}{3\ka^4}},\cH_{\Lambda}\right\}\sim0\,,
\ee
the equality holds on-shell,  assuming that the Hamiltonian constraint is satisfied, $\cH_{\Lambda}=0$.

\subsection{Physical trajectories from $\SL(2,\mathbb{R})$ flow}

We can compute the evolution in proper time of $v$ and $\cC$ from the Hamiltonian flow generated by the scalar constraint $\cH_{\Lambda}$ and integrated  in terms of $\SL(2,\R)$ group elements, as explained in section \ref{evolution} for the case $\Lambda=0$. For $\Lambda<0$, this leads to:
\begin{align}
\cH_{\Lambda}[\tau]&=\cH_{\Lambda}^{(0)}\,, \\
\cC[\tau]&=
\cC^{(0)}\cosh
-\f1{\ka^{2}}\sqrt{\f{-2\Lambda}3}\,\left(v^{(0)}-\f{3\ka^{2}}{2\Lambda}\cH_{\Lambda}^{(0)}\right)\,\sinh
\,,\nn\\
v[\tau]&=
\f{3\ka^{2}}{2\Lambda}\cH_{\Lambda}^{(0)}\left(1-\cosh\right)
+v^{(0)}\cosh
-\ka^{2}\cC^{(0)}\sqrt{\f3{-2\Lambda}}\sinh
\,,\nn
\end{align}
where $\cosh$ and $\sinh$ respectively stand for $\cosh[\tau\sqrt{-2\Lambda/3}]$ and $\sinh[\tau\sqrt{-2\Lambda/3}]$, 
and similarly with trigonometric functions for $\Lambda >0$. These expressions are of course linear in the initial conditions $v^{(0)}$, $\cC^{(0)}$ and $\cH^{(0)}$. Inverting them to get the initial conditions in terms of $v$, $\cC$ and $\cH$ at arbitrary time will give time-dependent conserved charges, which should generate the $\SL(2,\R)$ symmetry according to Noether. We postpone to future study the derivation of the corresponding precise action of Mobius transformations in the case with a non-vanishing cosmological constant at the level of the action (\ref{lambdaaction}). It is nevertheless interesting to notice that, according to the map to conformal mechanics,  the modified scalar constraint \eqref{Hlambda} with $\Lambda \ne 0$ corresponds to a modified potential with an extra-term in $q^2$ besides the conformal potential term in $q^{-2}$.

\section{$\SL(2,\R)$ Unitary Representations}
\label{su11irrep}
\label{sec:sl2rirrep}

At the quantum level, the commutators of the  $\sl(2,\R)\sim\su(1,1)$ Lie algebra:
\be
[\hj_{z},\hj_{\pm}]=\pm \hk_{\pm}
\,,\quad
[\hk_{+},\hk_{-}]=-2\hj_{z}
\,.
\ee
These commutators can be represented  as acting on states $|j,m\ra$, where the spin $j\in\C$ gives the value of the Casimir $\mathfrak{C}=\hj_{z}^{2}-\f12(\hk_{-}\hk_{+}+\hk_{+}\hk_{-})$, as:
\begin{align}
\hj_z \,| j, m \ra &= m \,| j, m \ra \\
\hk_{+} \,| j, m \ra &= \sqrt{m(m+1) -  j(j+1)}\, | j, m +1 \ra \\
\hk_{-}\, | j, m \ra &= \sqrt{m(m-1) -  j(j+1)}\, | j, m - 1 \ra \\
\what{\mathfrak{C}} \,| j, m \ra &= j(j+1)\, | j, m \ra
\end{align}
The unitary (irreducible) representations of the $\sl(2,\R)$ Lie algebra are obtained by making sure that $\hk_{+}^\dagger=\hk_{-}$.
This leads to  three classes of unitary representations:
\begin{itemize}
\item {\it The discrete series}:
We distinguish the positive and negative series, which consist respectively in lowest and highest weight representations. They are labelled by half-integers $j\in\f\N2$:
\be
\cD^{+}_j =\bigoplus_{m\in j+1+\N}\C\,| j, m \ra
\,,\quad
 \cD^{-}_j =\bigoplus_{m\in -j-1-\N}\C\,| j, m \ra
\ee
The Casimir is given by $\mathfrak{C} = j (j+1) \geqslant 0$. These can be constructed from the canonical quantization of the pair of harmonic oscillators in the spinorial formulation  used in \cite{Livine:2012mh} to define cosmological coherent states.

\item {\it The principal continuous series}: they are labelled by a real positive number $s\in\R^+$ and a parity $\eps=\pm$. Even parity representations are spanned by  states with integer magnetic moment $m\in\Z$ while odd parity representations are spanned by half-integer states $m\in\Z+\f12$:
\be
P^{+}_s= \bigoplus_{m\in \Z}\C\,| j, m \ra
\,,\quad
P^{-}_s= \bigoplus_{m\in \Z+\f12}\C\,| j, m \ra
\ee
The quadratic Casimir is now strictly negative, $\mathfrak{C} = - \left( s^2 + \frac{1}{4} \right) < 0$. It can be understood as $\mathfrak{C} = j(j+1)$ with $j=-\f12+is$. These are the $\SU(1,1)$ representations that we use in the present work to quantize the cosmological phase space. 

\item {\it The complementary series}: They are labelled by a real number $-\f12<j<0$ and interpolate between the discrete series starting at $j=0$ and the principal continuous series on the imaginary line $\mathfrak{Re}(j)=-\f12$:
\begin{align}
P^c_j= \bigoplus_{m\in \Z}\C\,| j, m \ra.
\end{align}
The Casimir is given by $\mathfrak{C}=j(j+1)$, with $- 1/4 < \mathfrak{C}  < 0$. These representations do not appear in the Plancherel decomposition for $L^2$ functions on $\SL(2,\R)$.

\end{itemize}

\section{Time overlap of thermal dilatation eigenstates}
\label{sec:boostedC}

In this section, we would like to consider eigenvectors of the ``boosted'' dilatation generator, $\hcC_{\beta}:=e^{+\beta \hcH}\hcC e^{-\beta \hcH}$ for an arbitrary real parameter $\beta\in\R$. Using the explicit formula  for the time-evolved operator $\hcC(\tau)=e^{+i\tau \hcH}\hcC e^{-i\tau \hcH}=\hcC-\tau\hcH$, derived from the $\SL(2,\R)$ flow generated by the CVH operators, we easily get:
\be
\hcC_{\beta}=e^{+\beta \hcH}\hcC e^{-\beta \hcH}
=\hcC+i\beta \hcH
\,,\qquad
\hcC_{\beta}^{\dagger}=\hcC_{-\beta}
\,.
\ee
This allows to compute the behavior of the time overlap $g(\tau)=\la \vphi^{\Delta}|e^{i\tau \hcH}| \vphi^{\Delta}\ra$ for an eigenstate $\hcC_{\beta}\,|\vphi^{\Delta}\ra=\Delta\,|\vphi^{\Delta}\ra$:
\beq
\tau \pp_{\tau}g
&=&
i\la \vphi| \tau \hcH e^{i\tau \hcH} |\vphi \ra 
=
i\la \vphi| \hcC e^{i\tau \hcH} |\vphi \ra -i\la \vphi| e^{i\tau \hcH}\hcC  |\vphi \ra 
\nn\\
&=&
i\la \vphi| (\hcC_{-\beta}+i\beta\hcH)\, e^{i\tau \hcH} |\vphi \ra -i\la \vphi| e^{i\tau \hcH}\, (\hcC_{\beta}-i\beta\hcH)  |\vphi \ra 
\nn\\
&=&
2\mathfrak{Im}(\Delta) g +2i\beta\pp_{\tau}g
\,,
\eeq
which implies the scaling in time, $g(\tau)\propto (\tau-2i\beta)^{2\mathfrak{Im}(\Delta)}$.
Notice that the computation of this overlap does not require the choice of a highest or lowest weight vector, as the two-point function computed in the body of the paper, which involves the volume operator $\hv$.




\begin{thebibliography}{ab}

\bibitem{Bekenstein:1973ur}
J.~D. Bekenstein, ``{Black holes and entropy},'' Phys. Rev. {\bf D7} (1973)
2333--2346.

\bibitem{Bardeen:1973gs}
J.~M. Bardeen, B.~Carter, and S.~W. Hawking, ``{The Four laws of black hole
  mechanics},'' Commun. Math. Phys. {\bf 31} (1973)
161--170.

\bibitem{Hawking:1974rv}
S.~W. Hawking, ``{Black hole explosions},'' Nature {\bf 248} (1974)
30--31.

\bibitem{Jacobson:1995ab}
T.~Jacobson, ``{Thermodynamics of space-time: The Einstein equation of
  state},'' Phys. Rev. Lett. {\bf 75} (1995) 1260--1263,
\href{http://arXiv.org/abs/gr-qc/9504004}{{\texttt{arXiv:gr-qc/9504004}}}.

\bibitem{Gibbons:1977mu}
G.~W. Gibbons and S.~W. Hawking, ``{Cosmological Event Horizons,
  Thermodynamics, and Particle Creation},'' Phys. Rev. {\bf D15} (1977)
2738--2751.

\bibitem{Frolov:2002va}
A.~V. Frolov and L.~Kofman, ``{Inflation and de Sitter thermodynamics},'' JCAP
  {\bf 0305} (2003) 009,
\href{http://arXiv.org/abs/hep-th/0212327}{{\texttt{arXiv:hep-th/0212327}}}.

\bibitem{Davis:2003ye}
T.~M. Davis, P.~C.~W. Davies, and C.~H. Lineweaver, ``{Black hole versus
  cosmological horizon entropy},'' Class. Quant. Grav. {\bf 20} (2003)
  2753--2764,
\href{http://arXiv.org/abs/astro-ph/0305121}{{\texttt{arXiv:astro-ph/0305121}}}.

\bibitem{Cai:2005ra}
R.-G. Cai and S.~P. Kim, ``{First law of thermodynamics and Friedmann equations
  of Friedmann-Robertson-Walker universe},'' JHEP {\bf 02} (2005) 050,
\href{http://arXiv.org/abs/hep-th/0501055}{{\texttt{arXiv:hep-th/0501055}}}.

\bibitem{Akbar:2006kj}
M.~Akbar and R.-G. Cai, ``{Thermodynamic Behavior of Friedmann Equations at
  Apparent Horizon of FRW Universe},'' Phys. Rev. {\bf D75} (2007) 084003,
\href{http://arXiv.org/abs/hep-th/0609128}{{\texttt{arXiv:hep-th/0609128}}}.

\bibitem{Cai:2008gw}
R.-G. Cai, L.-M. Cao, and Y.-P. Hu, ``{Hawking Radiation of Apparent Horizon in
  a FRW Universe},'' Class. Quant. Grav. {\bf 26} (2009) 155018,
\href{http://arXiv.org/abs/0809.1554}{{\texttt{arXiv:0809.1554}}}.

\bibitem{Carlip:1998wz}
S.~Carlip, ``{Black hole entropy from conformal field theory in any
  dimension},'' Phys. Rev. Lett. {\bf 82} (1999) 2828--2831,
\href{http://arXiv.org/abs/hep-th/9812013}{{\texttt{arXiv:hep-th/9812013}}}.

\bibitem{Carlip:1999cy}
S.~Carlip, ``{Entropy from conformal field theory at Killing horizons},''
  Class. Quant. Grav. {\bf 16} (1999) 3327--3348,
\href{http://arXiv.org/abs/gr-qc/9906126}{{\texttt{arXiv:gr-qc/9906126}}}.

\bibitem{Carlip:2002be}
S.~Carlip, ``{Near horizon conformal symmetry and black hole entropy},'' Phys.
  Rev. Lett. {\bf 88} (2002) 241301,
\href{http://arXiv.org/abs/gr-qc/0203001}{{\texttt{arXiv:gr-qc/0203001}}}.

\bibitem{Carlip:2011vr}
S.~Carlip, ``{Effective Conformal Descriptions of Black Hole Entropy},''
  Entropy {\bf 13} (2011) 1355--1379,
\href{http://arXiv.org/abs/1107.2678}{{\texttt{arXiv:1107.2678}}}.

\bibitem{Carlip:2012ff}
S.~Carlip, ``{Effective Conformal Descriptions of Black Hole Entropy: A
  Review},'' AIP Conf. Proc. {\bf 1483} (2012), no.~1, 54--62,
\href{http://arXiv.org/abs/1207.1488}{{\texttt{arXiv:1207.1488}}}.

\bibitem{Bardeen:1999px}
J.~M. Bardeen and G.~T. Horowitz, ``{The Extreme Kerr throat geometry: A Vacuum
  analog of AdS(2) x S**2},'' Phys. Rev. {\bf D60} (1999) 104030,
\href{http://arXiv.org/abs/hep-th/9905099}{{\texttt{arXiv:hep-th/9905099}}}.

\bibitem{Kunduri:2007vf}
H.~K. Kunduri, J.~Lucietti, and H.~S. Reall, ``{Near-horizon symmetries of
  extremal black holes},'' Class. Quant. Grav. {\bf 24} (2007) 4169--4190,
\href{http://arXiv.org/abs/0705.4214}{{\texttt{arXiv:0705.4214}}}.

\bibitem{Kunduri:2013ana}
H.~K. Kunduri and J.~Lucietti, ``{Classification of near-horizon geometries of
  extremal black holes},'' Living Rev. Rel. {\bf 16} (2013) 8,
\href{http://arXiv.org/abs/1306.2517}{{\texttt{arXiv:1306.2517}}}.

\bibitem{Hajian:2013lna}
K.~Hajian, A.~Seraj, and M.~M. Sheikh-Jabbari, ``{NHEG Mechanics: Laws of Near
  Horizon Extremal Geometry (Thermo)Dynamics},'' JHEP {\bf 03} (2014) 014,
\href{http://arXiv.org/abs/1310.3727}{{\texttt{arXiv:1310.3727}}}.

\bibitem{Compere:2015mza}
G.~Comp\`ere, K.~Hajian, A.~Seraj, and M.~M. Sheikh-Jabbari, ``{Extremal
  Rotating Black Holes in the Near-Horizon Limit: Phase Space and Symmetry
  Algebra},'' Phys. Lett. {\bf B749} (2015) 443--447,
  \href{http://arXiv.org/abs/1503.07861}{{\texttt{arXiv:1503.07861}}}.
[Phys. Lett.B749,443(2015)].

\bibitem{Maldacena:1997ih}
J.~M. Maldacena and A.~Strominger, ``{Universal low-energy dynamics for
  rotating black holes},'' Phys. Rev. {\bf D56} (1997) 4975--4983,
\href{http://arXiv.org/abs/hep-th/9702015}{{\texttt{arXiv:hep-th/9702015}}}.

\bibitem{Bredberg:2009pv}
I.~Bredberg, T.~Hartman, W.~Song, and A.~Strominger, ``{Black Hole
  Superradiance From Kerr/CFT},'' JHEP {\bf 04} (2010) 019,
\href{http://arXiv.org/abs/0907.3477}{{\texttt{arXiv:0907.3477}}}.

\bibitem{Birmingham:2001pj}
D.~Birmingham, I.~Sachs, and S.~N. Solodukhin, ``{Conformal field theory
  interpretation of black hole quasinormal modes},'' Phys. Rev. Lett. {\bf 88}
  (2002) 151301,
\href{http://arXiv.org/abs/hep-th/0112055}{{\texttt{arXiv:hep-th/0112055}}}.

\bibitem{Chen:2010sn}
B.~Chen and J.-j. Zhang, ``{Quasi-normal Modes of Extremal Black Holes from
  Hidden Conformal Symmetry},'' Phys. Lett. {\bf B699} (2011) 204--210,
\href{http://arXiv.org/abs/1012.2219}{{\texttt{arXiv:1012.2219}}}.

\bibitem{Chen:2010ik}
B.~Chen and J.~Long, ``{Hidden Conformal Symmetry and Quasi-normal Modes},''
  Phys. Rev. {\bf D82} (2010) 126013,
\href{http://arXiv.org/abs/1009.1010}{{\texttt{arXiv:1009.1010}}}.

\bibitem{Castro:2010fd}
A.~Castro, A.~Maloney, and A.~Strominger, ``{Hidden Conformal Symmetry of the
  Kerr Black Hole},'' Phys. Rev. {\bf D82} (2010) 024008,
\href{http://arXiv.org/abs/1004.0996}{{\texttt{arXiv:1004.0996}}}.

\bibitem{Bertini:2011ga}
S.~Bertini, S.~L. Cacciatori, and D.~Klemm, ``{Conformal structure of the
  Schwarzschild black hole},'' Phys. Rev. {\bf D85} (2012) 064018,
\href{http://arXiv.org/abs/1106.0999}{{\texttt{arXiv:1106.0999}}}.

\bibitem{Lowe:2011aa}
D.~A. Lowe and A.~Skanata, ``{Generalized Hidden Kerr/CFT},'' J. Phys. {\bf
  A45} (2012) 475401,
\href{http://arXiv.org/abs/1112.1431}{{\texttt{arXiv:1112.1431}}}.

\bibitem{Porfyriadis:2014fja}
A.~P. Porfyriadis and A.~Strominger, ``{Gravity waves from the Kerr/CFT
  correspondence},'' Phys. Rev. {\bf D90} (2014), no.~4, 044038,
\href{http://arXiv.org/abs/1401.3746}{{\texttt{arXiv:1401.3746}}}.

\bibitem{Pathak:2016vfc}
A.~Pathak, A.~P. Porfyriadis, A.~Strominger, and O.~Varela, ``{Logarithmic
  corrections to black hole entropy from Kerr/CFT},'' JHEP {\bf 04} (2017) 090,
\href{http://arXiv.org/abs/1612.04833}{{\texttt{arXiv:1612.04833}}}.

\bibitem{Guica:2008mu}
M.~Guica, T.~Hartman, W.~Song, and A.~Strominger, ``{The Kerr/CFT
  Correspondence},'' Phys. Rev. {\bf D80} (2009) 124008,
\href{http://arXiv.org/abs/0809.4266}{{\texttt{arXiv:0809.4266}}}.

\bibitem{Bredberg:2011hp}
I.~Bredberg, C.~Keeler, V.~Lysov, and A.~Strominger, ``{Cargese Lectures on the
  Kerr/CFT Correspondence},'' Nucl. Phys. Proc. Suppl. {\bf 216} (2011)
  194--210,
\href{http://arXiv.org/abs/1103.2355}{{\texttt{arXiv:1103.2355}}}.

\bibitem{Compere:2012jk}
G.~Comp\`ere, ``{The Kerr/CFT correspondence and its extensions},'' Living Rev.
  Rel. {\bf 15} (2012) 11,
  \href{http://arXiv.org/abs/1203.3561}{{\texttt{arXiv:1203.3561}}}.
[Living Rev. Rel.20,no.1,1(2017)].

\bibitem{Anninos:2011af}
D.~Anninos, S.~A. Hartnoll, and D.~M. Hofman, ``{Static Patch Solipsism:
  Conformal Symmetry of the de Sitter Worldline},'' Class. Quant. Grav. {\bf
  29} (2012) 075002,
\href{http://arXiv.org/abs/1109.4942}{{\texttt{arXiv:1109.4942}}}.
  
  \bibitem{Kehagias:2013xga} 
  A.~Kehagias and A.~Riotto,
  ``{Conformal Symmetries of FRW Accelerating Cosmologies,}''
  Nucl.\ Phys.\ B {\bf 884}, 547 (2014)
  doi:10.1016/j.nuclphysb.2014.05.006
  [arXiv:1309.3671 [hep-th]].
  
    \bibitem{Kehagias:2012pd} 
  A.~Kehagias and A.~Riotto,
  ``{Operator Product Expansion of Inflationary Correlators and Conformal Symmetry of de Sitter,}''
  Nucl.\ Phys.\ B {\bf 864}, 492 (2012)
  doi:10.1016/j.nuclphysb.2012.07.004
  [arXiv:1205.1523 [hep-th]].
  
  \bibitem{Kehagias:2015jha} 
  A.~Kehagias and A.~Riotto,
  ``{High Energy Physics Signatures from Inflation and Conformal Symmetry of de Sitter,}''
  Fortsch.\ Phys.\  {\bf 63}, 531 (2015)
  doi:10.1002/prop.201500025
  [arXiv:1501.03515 [hep-th]].
 
  
  \bibitem{Kehagias:2017rpe} 
  A.~Kehagias and A.~Riotto,
  ``{Inflation and Conformal Invariance: The Perspective from Radial Quantization,}''
  Fortsch.\ Phys.\  {\bf 65}, no. 5, 1700023 (2017)
  doi:10.1002/prop.201700023
  [arXiv:1701.05462 [hep-th]].
  
  \bibitem{Baumann:2019oyu} 
  D.~Baumann, C.~D.~Pueyo, A.~Joyce, H.~Lee and G.~L.~Pimentel,
  ``{The Cosmological Bootstrap: Weight-Shifting Operators and Scalar Seeds,}''
  arXiv:1910.14051 [hep-th].
  
  \bibitem{Arkani-Hamed:2018kmz} 
  N.~Arkani-Hamed, D.~Baumann, H.~Lee and G.~L.~Pimentel,
  ``{The Cosmological Bootstrap: Inflationary Correlators from Symmetries and Singularities,}''
  arXiv:1811.00024 [hep-th]

\bibitem{Strominger:2001pn}
A.~Strominger, ``{The dS / CFT correspondence},'' JHEP {\bf 10} (2001) 034,
\href{http://arXiv.org/abs/hep-th/0106113}{{\texttt{arXiv:hep-th/0106113}}}.

\bibitem{Dyson:2002nt}
L.~Dyson, J.~Lindesay, and L.~Susskind, ``{Is there really a de Sitter/CFT
  duality?},'' JHEP {\bf 08} (2002) 045,
\href{http://arXiv.org/abs/hep-th/0202163}{{\texttt{arXiv:hep-th/0202163}}}.

\bibitem{Anninos:2011ui}
D.~Anninos, T.~Hartman, and A.~Strominger, ``{Higher Spin Realization of the
  dS/CFT Correspondence},'' Class. Quant. Grav. {\bf 34} (2017), no.~1, 015009,
\href{http://arXiv.org/abs/1108.5735}{{\texttt{arXiv:1108.5735}}}.

\bibitem{Compere:2014cna}
G.~Comp\`ere, L.~Donnay, P.-H. Lambert, and W.~Schulgin, ``{Liouville theory
  beyond the cosmological horizon},'' JHEP {\bf 03} (2015) 158,
\href{http://arXiv.org/abs/1411.7873}{{\texttt{arXiv:1411.7873}}}.

\bibitem{Neiman:2017zdr}
Y.~Neiman, ``{Towards causal patch physics in dS/CFT},'' EPJ Web Conf. {\bf
  168} (2018) 01007,
\href{http://arXiv.org/abs/1710.05682}{{\texttt{arXiv:1710.05682}}}.

\bibitem{Donnay:2019zif}
L.~Donnay and G.~Giribet, ``{Cosmological horizons, Noether charges and
  entropy},'' Class. Quant. Grav. {\bf 36} (2019), no.~16, 165005,
\href{http://arXiv.org/abs/1903.09271}{{\texttt{arXiv:1903.09271}}}.

\bibitem{Grumiller:2019fmp} 
  D.~Grumiller, A.~Pérez, M.~M.~Sheikh-Jabbari, R.~Troncoso and C.~Zwikel,
  ``{Spacetime structure near generic horizons and soft hair,}''
  \href{http://arXiv.org/abs/1908.09833}{{\texttt{arXiv:1908.09833}}}.
  
  \bibitem{Hartle:1983ai} 
  J.~B.~Hartle and S.~W.~Hawking,
  ``{Wave Function of the Universe,}''
  Phys.\ Rev.\ D {\bf 28}, 2960 (1983)
    \href{http://doi:10.1103/PhysRevD.28.2960}{{\texttt{doi:10.1103/PhysRevD.28.2960}}}.
  
  \bibitem{Vilenkin:1982de} 
  A.~Vilenkin,
  ``{Creation of Universes from Nothing,}''
  Phys.\ Lett.\  {\bf 117B}, 25 (1982).
      \href{http://doi:10.1016/0370-2693(82)90866-8}{{\texttt{doi:10.1016/0370-2693(82)90866-8}}}.
  
  
  \bibitem{Feldbrugge:2017kzv} 
  J.~Feldbrugge, J.~L.~Lehners and N.~Turok,
  ``{Lorentzian Quantum Cosmology,}''
  Phys.\ Rev.\ D {\bf 95}, no. 10, 103508 (2017)
  \href{http://arXiv.org/abs/1703.02076}{{\texttt{arXiv:1703.02076}}}.
  
  \bibitem{Vilenkin:2018dch} 
  A.~Vilenkin and M.~Yamada,
  ``{Tunneling wave function of the universe,}''
  Phys.\ Rev.\ D {\bf 98}, no. 6, 066003 (2018)
    \href{http://arXiv.org/abs/1808.02032}{{\texttt{arXiv:1808.02032}}}.

\bibitem{Bojowald:2015iga}
M.~Bojowald, ``{Quantum cosmology: a review},'' Rept. Prog. Phys. {\bf 78}
  (2015) 023901,
\href{http://arXiv.org/abs/1501.04899}{{\texttt{arXiv:1501.04899}}}.

\bibitem{Bojowald:2012xy}
M.~Bojowald, ``{Quantum Cosmology: Effective Theory},'' Class. Quant. Grav.
  {\bf 29} (2012) 213001,
\href{http://arXiv.org/abs/1209.3403}{{\texttt{arXiv:1209.3403}}}.

\bibitem{Bojowald:2010cj}
M.~Bojowald, C.~Kiefer, and P.~Vargas~Moniz, ``{Quantum cosmology for the 21st
  century: A Debate},'' in {\em {On recent developments in theoretical and
  experimental general relativity, astrophysics and relativistic field
  theories. Proceedings, 12th Marcel Grossmann Meeting on General Relativity,
  Paris, France, July 12-18, 2009. Vol. 1-3}}, pp.~589--608.
\newblock 2010.
\newblock
\href{http://arXiv.org/abs/1005.2471}{{\texttt{arXiv:1005.2471}}}.
\newblock

\bibitem{Halliwell:2002cg}
J.~J. Halliwell, ``{The Interpretation of quantum cosmology and the problem of
  time},'' in {\em {The future of theoretical physics and cosmology:
  Celebrating Stephen Hawking's 60th birthday. Proceedings, Workshop and
  Symposium, Cambridge, UK, January 7-10, 2002}}, pp.~675--692.
\newblock 2002.
\newblock
\href{http://arXiv.org/abs/gr-qc/0208018}{{\texttt{arXiv:gr-qc/0208018}}}.
\newblock

\bibitem{Vilenkin:1994ua}
A.~Vilenkin, ``{Predictions from quantum cosmology},'' Phys. Rev. Lett. {\bf
  74} (1995) 846--849,
\href{http://arXiv.org/abs/gr-qc/9406010}{{\texttt{arXiv:gr-qc/9406010}}}.

\bibitem{Hartle:1997hw}
J.~B. Hartle, ``{Quantum cosmology: Problems for the 21st century},'' in {\em
  {Physics in the 21st century. Proceedings, 11th Nishinomiya-Yukawa Memorial
  Symposium, Nishinomiya, Japan, November 7-8, 1996}}, pp.~179--199.
\newblock 1997.
\newblock
\href{http://arXiv.org/abs/gr-qc/9701022}{{\texttt{arXiv:gr-qc/9701022}}}.
\newblock

\bibitem{BenAchour:2019ywl}
  J.~Ben Achour and E.~R.~Livine,
  ``{Protected $SL(2,\mathbb{R})$ Symmetry in Quantum Cosmology,}''
  JCAP {\bf 1909}, 012 (2019)
  \href{http://arXiv.org/abs/1904.06149}{{\texttt{arXiv:1904.06149}}}.
  
  \bibitem{BenAchour:2018jwq} 
  J.~Ben Achour and E.~R.~Livine,
  ``{Polymer Quantum Cosmology: Lifting quantization ambiguities using a $SL(2,\mathbb{R})$ conformal symmetry,}''
  Phys.\ Rev.\ D {\bf 99}, no. 12, 126013 (2019)
  \href{http://arXiv.org/abs/1806.09290}{{\texttt{arXiv:1806.09290}}}.

\bibitem{BenAchour:2017qpb}
J.~Ben~Achour and E.~R. Livine, ``{Thiemann complexifier in classical and
  quantum FLRW cosmology},'' Phys. Rev. {\bf D96} (2017), no.~6, 066025,
\href{http://arXiv.org/abs/1705.03772}{{\texttt{arXiv:1705.03772}}}.


\bibitem{Bodendorfer:2019wik}   
  N.~Bodendorfer and D.~Wuhrer,
  ``{Renormalisation with SU(1, 1) coherent states on the LQC Hilbert space,}''
  \href{http://arXiv.org/abs/1904.13269}{{\texttt{arXiv:1904.13269}}}.
  
  \bibitem{Bodendorfer:2018csn} 
  N.~Bodendorfer and F.~Haneder,
  ``{Coarse graining as a representation change,}''
  Phys.\ Lett.\ B {\bf 792}, 69 (2019)
    \href{http://arXiv.org/abs/1811.02792}{{\texttt{arXiv:1811.02792}}}.
    
    \bibitem{Livine:2012mh} 
  E.~R.~Livine and M.~Martin-Benito,
  ``{Group theoretical Quantization of Isotropic Loop Cosmology,}''
  Phys.\ Rev.\ D {\bf 85}, 124052 (2012)
  \    \href{http://arXiv.org/abs/1204.0539}{{\texttt{arXiv:1204.0539}}}.
  
  \bibitem{Bojowald:2007bg} 
    M.~Bojowald,
  ``{Dynamical coherent states and physical solutions of quantum cosmological bounces,}''
  Phys.\ Rev.\ D {\bf 75}, 123512 (2007)
    \    \href{http://arXiv.org/abs/0703144}{{\texttt{arXiv:0703144}}}.

\bibitem{deAlfaro:1976vlx}
V.~de~Alfaro, S.~Fubini, and G.~Furlan, ``{Conformal Invariance in Quantum
  Mechanics},'' Nuovo Cim. {\bf A34} (1976)
569.

\bibitem{Andrzejewski:2011ya}
K.~Andrzejewski and J.~Gonera, ``{On the geometry of conformal mechanics},''
\href{http://arXiv.org/abs/1108.1299}{{\texttt{arXiv:1108.1299}}}.

\bibitem{Andrzejewski:2015jya}
K.~Andrzejewski, ``{Quantum conformal mechanics emerging from unitary
  representations of SL(2,$\mathbb{R}$)},'' Annals Phys. {\bf 367} (2016)
  227--250,
\href{http://arXiv.org/abs/1506.05596}{{\texttt{arXiv:1506.05596}}}.

\bibitem{Okazaki:2017lpn}
T.~Okazaki, ``{Implications of Conformal Symmetry in Quantum Mechanics},''
  Phys. Rev. {\bf D96} (2017), no.~6, 066030,
\href{http://arXiv.org/abs/1704.00286}{{\texttt{arXiv:1704.00286}}}.

\bibitem{Cadoni:2000iz}
M.~Cadoni, P.~Carta, and S.~Mignemi, ``{A Realization of the
  infinite-dimensional symmetries of conformal mechanics},'' Phys. Rev. {\bf
  D62} (2000) 086002,
\href{http://arXiv.org/abs/hep-th/0004107}{{\texttt{arXiv:hep-th/0004107}}}.

\bibitem{Khodaee:2017tbk}
S.~Khodaee and D.~Vassilevich, ``{Note on correlation functions in conformal
  quantum mechanics},'' Mod. Phys. Lett. {\bf A32} (2017), no.~29, 1750157,
\href{http://arXiv.org/abs/1706.10225}{{\texttt{arXiv:1706.10225}}}.

\bibitem{Carinena:2017zfy}
J.~F. Cari\~nena, L.~Inzunza, and M.~S. Plyushchay, ``{Rational deformations of
  conformal mechanics},'' Phys. Rev. {\bf D98} (2018), no.~2, 026017,
\href{http://arXiv.org/abs/1707.07357}{{\texttt{arXiv:1707.07357}}}.

\bibitem{Mignemi:2001uz}
S.~Mignemi, ``{Black holes and conformal mechanics},'' Mod. Phys. Lett. {\bf
  A16} (2001) 1997--2002,
\href{http://arXiv.org/abs/hep-th/0104175}{{\texttt{arXiv:hep-th/0104175}}}.

\bibitem{Camblong:2003mz}
H.~E. Camblong and C.~R. Ordonez, ``{Anomaly in conformal quantum mechanics:
  From molecular physics to black holes},'' Phys. Rev. {\bf D68} (2003) 125013,
\href{http://arXiv.org/abs/hep-th/0303166}{{\texttt{arXiv:hep-th/0303166}}}.

\bibitem{Clement:2001ny}
G.~Clement and D.~Gal'tsov, ``{Conformal mechanics on rotating
  Bertotti-Robinson space-time},'' Nucl. Phys. {\bf B619} (2001) 741--759,
\href{http://arXiv.org/abs/hep-th/0105237}{{\texttt{arXiv:hep-th/0105237}}}.

\bibitem{Gaiotto:2004ij}
D.~Gaiotto, A.~Strominger, and X.~Yin, ``{Superconformal black hole quantum
  mechanics},'' JHEP {\bf 11} (2005) 017,
\href{http://arXiv.org/abs/hep-th/0412322}{{\texttt{arXiv:hep-th/0412322}}}.

\bibitem{Camblong:2004ye}
H.~E. Camblong and C.~R. Ordonez, ``{Black hole thermodynamics from
  near-horizon conformal quantum mechanics},'' Phys. Rev. {\bf D71} (2005)
  104029,
\href{http://arXiv.org/abs/hep-th/0411008}{{\texttt{arXiv:hep-th/0411008}}}.

\bibitem{Strominger:2003tm}
A.~Strominger, ``{A Matrix model for AdS(2)},'' JHEP {\bf 03} (2004) 066,
\href{http://arXiv.org/abs/hep-th/0312194}{{\texttt{arXiv:hep-th/0312194}}}.

\bibitem{Spradlin:1999bn}
M.~Spradlin and A.~Strominger, ``{Vacuum states for AdS(2) black holes},'' JHEP
  {\bf 11} (1999) 021,
\href{http://arXiv.org/abs/hep-th/9904143}{{\texttt{arXiv:hep-th/9904143}}}.

\bibitem{Azeyanagi:2007bj}
T.~Azeyanagi, T.~Nishioka, and T.~Takayanagi, ``{Near Extremal Black Hole
  Entropy as Entanglement Entropy via AdS(2)/CFT(1)},'' Phys. Rev. {\bf D77}
  (2008) 064005,
\href{http://arXiv.org/abs/0710.2956}{{\texttt{arXiv:0710.2956}}}.

\bibitem{Hartman:2008dq}
T.~Hartman and A.~Strominger, ``{Central Charge for AdS(2) Quantum Gravity},''
  JHEP {\bf 04} (2009) 026,
\href{http://arXiv.org/abs/0803.3621}{{\texttt{arXiv:0803.3621}}}.

\bibitem{Chamon:2011xk}
C.~Chamon, R.~Jackiw, S.-Y. Pi, and L.~Santos, ``{Conformal quantum mechanics
  as the CFT$_1$ dual to AdS$_2$},'' Phys. Lett. {\bf B701} (2011) 503--507,
\href{http://arXiv.org/abs/1106.0726}{{\texttt{arXiv:1106.0726}}}.

\bibitem{Axenides:2013iwa}
M.~Axenides, E.~G. Floratos, and S.~Nicolis, ``{Modular discretization of the
  AdS$_{2}$/CFT$_{1}$ holography},'' JHEP {\bf 02} (2014) 109,
\href{http://arXiv.org/abs/1306.5670}{{\texttt{arXiv:1306.5670}}}.

\bibitem{Pinzul:2017wch}
A.~Pinzul and A.~Stern, ``{Non-commutative $AdS_2/CFT_1$ duality: the case of
  massless scalar fields},'' Phys. Rev. {\bf D96} (2017), no.~6, 066019,
\href{http://arXiv.org/abs/1707.04816}{{\texttt{arXiv:1707.04816}}}.

\bibitem{Mezei:2017kmw}
M.~Mezei, S.~S. Pufu, and Y.~Wang, ``{A 2d/1d Holographic Duality},''
\href{http://arXiv.org/abs/1703.08749}{{\texttt{arXiv:1703.08749}}}.

\bibitem{Gupta:2017xex}
K.~S. Gupta, E.~Harikumar, and N.~S. Zuhair, ``{Conformal quantum mechanics and
  holography in noncommutative space?time},'' Phys. Lett. {\bf B772} (2017)
  808--813,
\href{http://arXiv.org/abs/1704.03666}{{\texttt{arXiv:1704.03666}}}.

\bibitem{Grumiller:2017qao}
D.~Grumiller, R.~McNees, J.~Salzer, C.~Valcárcel, and D.~Vassilevich,
  ``{Menagerie of AdS$_{2}$ boundary conditions},'' JHEP {\bf 10} (2017) 203,
\href{http://arXiv.org/abs/1708.08471}{{\texttt{arXiv:1708.08471}}}.

\bibitem{Kolekar:2018sba}
K.~S. Kolekar and K.~Narayan, ``{AdS$_2$ dilaton gravity from reductions of
  some nonrelativistic theories},'' Phys. Rev. {\bf D98} (2018), no.~4, 046012,
\href{http://arXiv.org/abs/1803.06827}{{\texttt{arXiv:1803.06827}}}.

\bibitem{Sarosi:2017ykf} 
  G.~Sárosi,
  ``{AdS$_{2}$ holography and the SYK model,}''
  PoS Modave {\bf 2017}, 001 (2018)
  \href{http://arXiv.org/abs/1711.08482}{{\texttt{arXiv:1711.08482}}}.

\bibitem{Poland:2018epd}
D.~Poland, S.~Rychkov, and A.~Vichi, ``{The Conformal Bootstrap: Theory,
  Numerical Techniques, and Applications},'' Rev. Mod. Phys. {\bf 91} (2019)
  015002,
\href{http://arXiv.org/abs/1805.04405}{{\texttt{arXiv:1805.04405}}}.

\bibitem{Kumar:1999fx}
J.~Kumar, ``{Conformal mechanics and the Virasoro algebra},'' JHEP {\bf 04}
  (1999) 006,
\href{http://arXiv.org/abs/hep-th/9901139}{{\texttt{arXiv:hep-th/9901139}}}.

\bibitem{Cacciatori:1999rp}
S.~Cacciatori, D.~Klemm, and D.~Zanon, ``{W(infinity) algebras, conformal
  mechanics, and black holes},'' Class. Quant. Grav. {\bf 17} (2000)
  1731--1748,
\href{http://arXiv.org/abs/hep-th/9910065}{{\texttt{arXiv:hep-th/9910065}}}.

\bibitem{Mignemi:2000cv}
S.~Mignemi, ``{A Note on the infinite dimensional symmetries of classical
  Hamiltonian systems},''
\href{http://arXiv.org/abs/hep-th/0004150}{{\texttt{arXiv:hep-th/0004150}}}.

\bibitem{Arzano:2018oby}
M.~Arzano and J.~Kowalski-Glikman, ``{Horizon temperature on the real line},''
  Phys. Lett. B {\bf C788} (2019) 82--86,
\href{http://arXiv.org/abs/1804.10550}{{\texttt{arXiv:1804.10550}}}.

\bibitem{AIHPA_1965__3_1_13_0}
L.~C. Biedenharn, J.~Nuyts, and N.~Straumann, ``On the unitary representations
  of $SU(1, 1)$ and $SU(2, 1)$,'' Annales de l'I.H.P. Physique th\'eorique {\bf
  3} (1965), no.~1, 13--39.

\bibitem{ruhl1970lorentz}
W.~R{\"u}hl, {\em The Lorentz group and harmonic analysis}.
\newblock Mathematical physics monograph series. W. A. Benjamin, 1970.

\bibitem{Kitaev:2017hnr}
A.~Kitaev, ``{Notes on $\widetilde{\mathrm{SL}}(2,\mathbb{R})$
  representations},''
\href{http://arXiv.org/abs/1711.08169}{{\texttt{arXiv:1711.08169}}}.

\bibitem{lindblad}
G.~Lindblad and B.~Nage, ``{Continuous bases for unitary irreducible
  representations of $SU(1, 1)$},'' Annales de l'I.H.P. Physique th\'eorique
  {\bf 13} (1970) 27--56.
  
  \bibitem{Jackiw:2012ur} 
  R.~Jackiw and S.-Y.~Pi,
  ``{Conformal Blocks for the 4-Point Function in Conformal Quantum Mechanics,}''
  Phys.\ Rev.\ D {\bf 86}, 045017 (2012)
  Erratum: [Phys.\ Rev.\ D {\bf 86}, 089905 (2012)]
  \href{http://arXiv.org/abs/1205.0443}{{\texttt{arXiv:1205.0443}}}.
  
  \bibitem{Hertog:2004rz} 
  T.~Hertog and G.~T.~Horowitz,
  ``{Towards a big crunch dual,}''
  JHEP {\bf 0407}, 073 (2004)
  \href{http://arXiv.org/abs/0406134}{{\texttt{arXiv:0406134}}}.
  
  \bibitem{Hertog:2005hu} 
  T.~Hertog and G.~T.~Horowitz,
  ``{Holographic description of AdS cosmologies,}''
  JHEP {\bf 0504}, 005 (2005)
  \href{http://arXiv.org/abs/0503071}{{\texttt{arXiv:0503071}}}.
  
  \bibitem{Craps:2007ch} 
  B.~Craps, T.~Hertog and N.~Turok,
  ``{On the Quantum Resolution of Cosmological Singularities using AdS/CFT,}''
  Phys.\ Rev.\ D {\bf 86}, 043513 (2012)
  \href{http://arXiv.org/abs/1712.4180}{{\texttt{arXiv:1712.4180}}}..

\bibitem{Gielen:2013naa}
S.~Gielen, D.~Oriti, and L.~Sindoni, ``{Homogeneous cosmologies as group field
  theory condensates},'' JHEP {\bf 06} (2014) 013,
\href{http://arXiv.org/abs/1311.1238}{{\texttt{arXiv:1311.1238}}}.

\bibitem{Oriti:2016ueo}
D.~Oriti, L.~Sindoni, and E.~Wilson-Ewing, ``{Bouncing cosmologies from quantum
  gravity condensates},'' Class. Quant. Grav. {\bf 34} (2017), no.~4, 04LT01,
\href{http://arXiv.org/abs/1602.08271}{{\texttt{arXiv:1602.08271}}}.

\bibitem{Oriti:2016qtz}
D.~Oriti, L.~Sindoni, and E.~Wilson-Ewing, ``{Emergent Friedmann dynamics with
  a quantum bounce from quantum gravity condensates},'' Class. Quant. Grav.
  {\bf 33} (2016), no.~22, 224001,
\href{http://arXiv.org/abs/1602.05881}{{\texttt{arXiv:1602.05881}}}.

\bibitem{Oriti:2016acw}
D.~Oriti, ``{The universe as a quantum gravity condensate},'' Comptes Rendus
  Physique {\bf 18} (2017) 235--245,
\href{http://arXiv.org/abs/1612.09521}{{\texttt{arXiv:1612.09521}}}.

\bibitem{Lidsey:2018byv}
J.~E. Lidsey, ``{Inflationary Cosmology, Diffeomorphism Group of the Line and
  Virasoro Coadjoint Orbits},''
\href{http://arXiv.org/abs/1802.09186}{{\texttt{arXiv:1802.09186}}}.

\bibitem{Mertens:2018fds}
T.~G. Mertens, ``{The Schwarzian theory - origins},'' JHEP {\bf 05} (2018) 036,
\href{http://arXiv.org/abs/1801.09605}{{\texttt{arXiv:1801.09605}}}.

\bibitem{Blommaert:2018oro}
A.~Blommaert, T.~G. Mertens, and H.~Verschelde, ``{The Schwarzian Theory - A
  Wilson Line Perspective},'' JHEP {\bf 12} (2018) 022,
\href{http://arXiv.org/abs/1806.07765}{{\texttt{arXiv:1806.07765}}}.

\bibitem{Mertens:2017mtv}
T.~G. Mertens, G.~J. Turiaci, and H.~L. Verlinde, ``{Solving the Schwarzian via
  the Conformal Bootstrap},'' JHEP {\bf 08} (2017) 136,
\href{http://arXiv.org/abs/1705.08408}{{\texttt{arXiv:1705.08408}}}.

\bibitem{Lam:2018pvp}
H.~T. Lam, T.~G. Mertens, G.~J. Turiaci, and H.~Verlinde, ``{Shockwave S-matrix
  from Schwarzian Quantum Mechanics},'' JHEP {\bf 11} (2018) 182,
\href{http://arXiv.org/abs/1804.09834}{{\texttt{arXiv:1804.09834}}}.

\bibitem{Belokurov:2018aol}
V.~V. Belokurov and E.~T. Shavgulidze, ``{Correlation functions in the
  Schwarzian theory},'' JHEP {\bf 11} (2018) 036,
\href{http://arXiv.org/abs/1804.00424}{{\texttt{arXiv:1804.00424}}}.

\bibitem{Haehl:2017pak}
F.~M. Haehl and M.~Rozali, ``{Fine Grained Chaos in $AdS_2$ Gravity},'' Phys.
  Rev. Lett. {\bf 120} (2018), no.~12, 121601,
\href{http://arXiv.org/abs/1712.04963}{{\texttt{arXiv:1712.04963}}}.

\bibitem{Jensen:2016pah}
K.~Jensen, ``{Chaos in AdS$_2$ Holography},'' Phys. Rev. Lett. {\bf 117}
  (2016), no.~11, 111601,
\href{http://arXiv.org/abs/1605.06098}{{\texttt{arXiv:1605.06098}}}.

\bibitem{Turiaci:2016cvo}
G.~Turiaci and H.~Verlinde, ``{On CFT and Quantum Chaos},'' JHEP {\bf 12}
  (2016) 110,
\href{http://arXiv.org/abs/1603.03020}{{\texttt{arXiv:1603.03020}}}.

\bibitem{Cornish:1997ah}
N.~J. Cornish and E.~P.~S. Shellard, ``{Chaos in quantum cosmology},'' Phys.
  Rev. Lett. {\bf 81} (1998) 3571--3574,
\href{http://arXiv.org/abs/gr-qc/9708046}{{\texttt{arXiv:gr-qc/9708046}}}.

\bibitem{Damour:2002et}
T.~Damour, M.~Henneaux, and H.~Nicolai, ``{Cosmological billiards},'' Class.
  Quant. Grav. {\bf 20} (2003) R145--R200,
\href{http://arXiv.org/abs/hep-th/0212256}{{\texttt{arXiv:hep-th/0212256}}}.



 \end{thebibliography}

\end{document}